\documentclass[epj]{svjour}
\usepackage{graphicx}
\usepackage{amssymb,amsmath} 
\usepackage{booktabs,units}

\providecommand{\be}{\begin{equation}}
\providecommand{\ee}{\end{equation}}
\providecommand{\sub}[1]{_{\rm #1}}

\providecommand{\bdma}{\begin{eqnarray*}}
\providecommand{\edma}{\end{eqnarray*}}
\providecommand{\bdm}{\begin{displaymath}}
\providecommand{\edm}{\end{displaymath}}
\providecommand{\ablpart}[2]{\frac{\partial #1}{\partial #2}} 
\providecommand{\bi}{\begin{itemize}}
\providecommand{\ei}{\end{itemize}}
\providecommand{\rhoQmax}{\rho_{\rm max}}  

\newcommand{\figpath}[1]{./#1}

\begin{document}
\title{Theoretical vs. Empirical Classification and Prediction of Congested Traffic States}
\author{Dirk Helbing,\inst{1} Martin Treiber,\inst{2} Arne Kesting\inst{2} \and Martin Sch\"onhof\inst{2}
}                     
%
%
\authorrunning{D. Helbing, M. Treiber, A. Kesting and M. Sch\"onhof}
\titlerunning{Classification and Prediction of Congested Traffic States}
\institute{$^{1}$ETH Zurich, UNO D11, Universit\"atstr. 41, 8092 Zurich, Switzerland\\
$^2$ Institute for Transport \& Economics, TU Dresden, Andreas-Schubert-Str. 23, 01062 Dresden, Germany}
\date{Received: date / Revised version: date}
%
\abstract{
Starting from the instability diagram of a traffic flow model, we
derive conditions for the occurrence of congested traffic states, their
appearance, their spreading in space and time, and the related
increase in travel times. We discuss the terminology of traffic phases
and give empirical evidence for the existence of a phase diagram of
traffic states. In contrast to previously presented phase diagrams, it
is shown that ``widening synchronized patterns''
are possible, if the maximum flow is located inside of a metastable density
regime. Moreover, for various kinds of traffic models with different
instability diagrams it is discussed, how the related phase diagrams
are expected to approximately look like. Apart from this, it is pointed out that combinations of
on- and off-ramps create different patterns than a
single, isolated on-ramp.
} 
\maketitle
\section{Introduction}
\begin{figure*}

\centering
 \includegraphics[width=1.02\textwidth]{\figpath{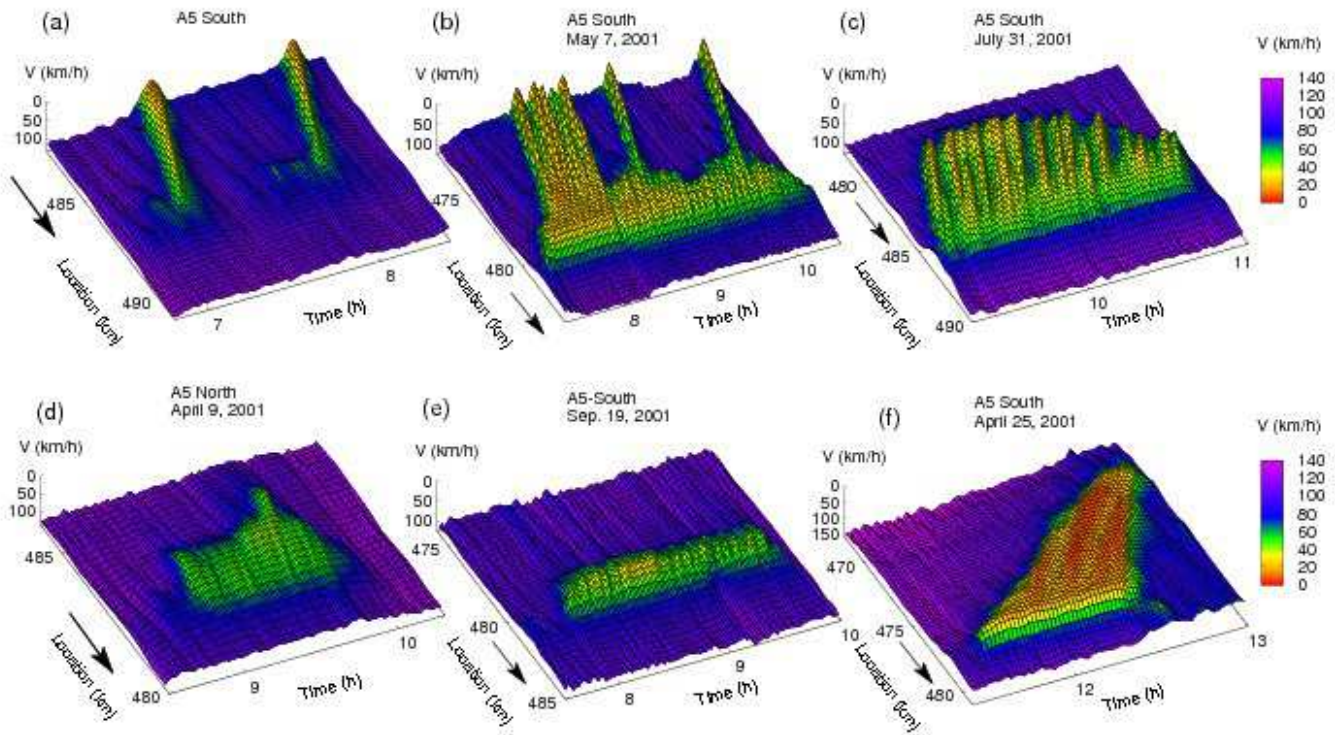}}

 \caption{\label{fig:elemPatterns}Examples of elementary
 patterns of congested traffic measured on the German freeway A5 close
 to Frankfurt. For better illustration of the traffic patterns, speeds are displayed upside down.
 The driving direction is indicated by arrows. 
 Top row: (a) Moving clusters (MC), (b) 
stop-and-go waves (SGW), (c) oscillating congested traffic (OCT). Bottom
row: (d) Widening synchronized pattern (WSP), (e) pinned localized cluster
(PLC), and (f) homogeneous congested traffic (HCT). The spatiotemporal
velocity fields have been reconstructed from one-minute data of double-loop detector
cross sections using the ``adaptive smoothing method''~protect\cite{Treiber-smooth}.}

\end{figure*}

\begin{figure}
\centering

 \includegraphics[width=0.5\textwidth]{\figpath{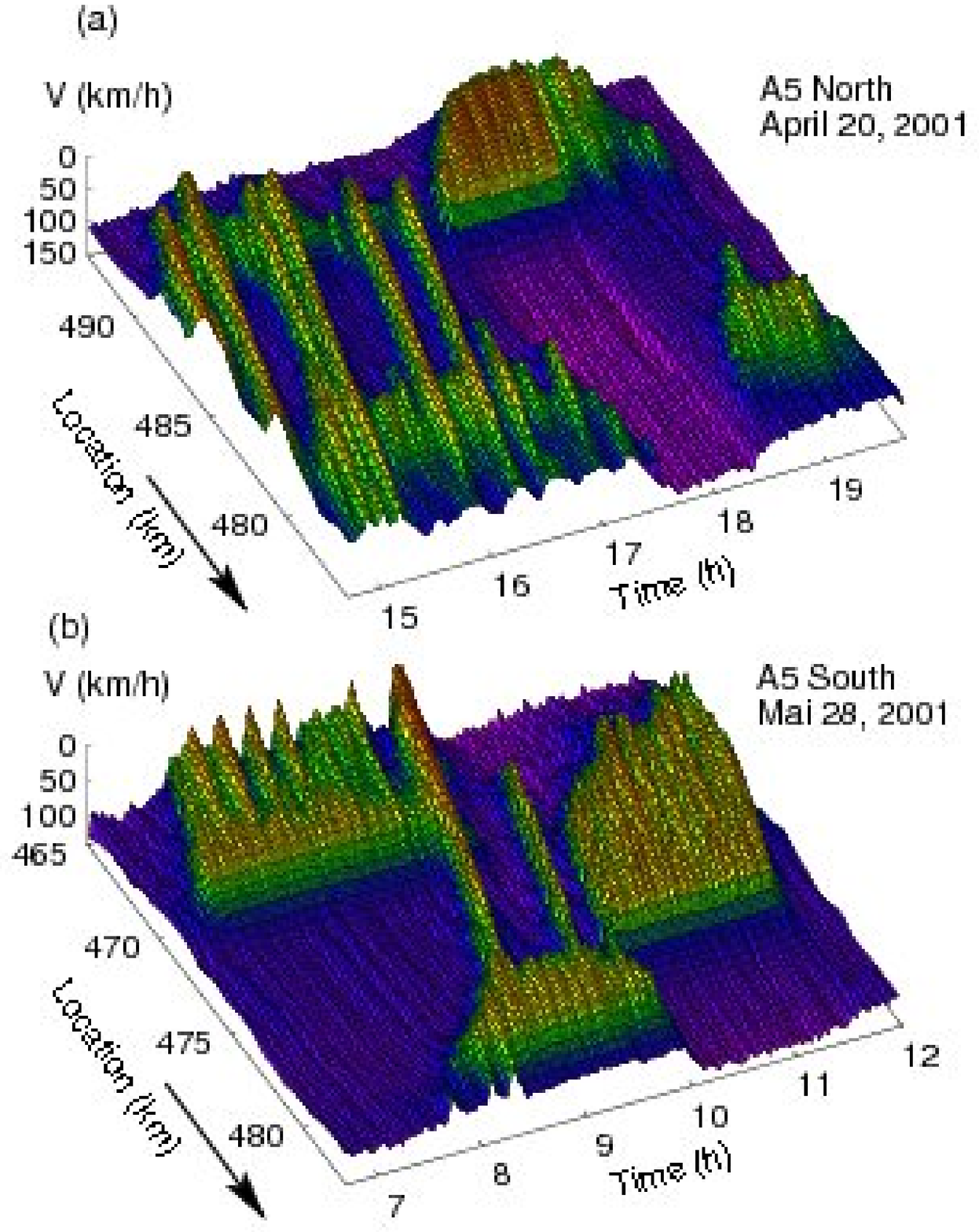}}

  \caption{\label{fig:combinedPatterns}Two examples of complex traffic
  states measured on the German freeway A5 close to Frankfurt. Top: On the A5 North,
  an accident occurs at $x=487.5$ km at the time $t= 17$:13~h, which
  causes a HCT pattern that turns into an OCT pattern as the upstream
traffic flow goes down.  
  The capacity drop related to the congestion pattern reduces the
downstream flow and leads to a dissolution of the previous SGW pattern
over there around $t=18$:00~h. Bottom: On the freeway A5 South, the
stop-and-go waves induced by a bottleneck at $x=480$ {km} replace the
OCT at the bottleneck near $x=470$~km. At time $t=9$:50~h, the waves
induce an accident at $x=478.33$~{km}, which triggers a new OCT
pattern further upstream. The related capacity drop, in turn, causes
the  previous OCT state at $x\approx 480$~km to dissolve.} 
\end{figure}

While traffic science makes a clear distinction between free and
congested traffic, the empirical analysis of spatiotemporal
congestion patterns has recently revealed an unexpected complexity of
traffic states. Early contributions in traffic physics focussed on the
study of so-called ``phantom traffic jams''~\cite{Treiterer},
i.e. traffic jams resulting from minor perturbations in the traffic
flow rather than from accidents, building sites, or other
bottlenecks. This subject has recently been revived due to new
technologies facilitating experimental traffic research
\cite{Sugiyama-NJP08}.  Related theoretical and numerical stability
analyses were~-- and still are~-- often carried out for setups with
periodic boundary conditions. This is, of course, quite artificial, as
compared to real traffic situations. Therefore, in response to
empirical findings \cite{Kerner-sync}, physicists have pointed out
that the occurrence of congested traffic on real freeways normally
results from a {\it combination} of three
ingredients~\cite{Phase,Arne-ACC-TRC}:
\begin{enumerate}
\item a high traffic volume (defined as the freeway flow plus the
actual on-ramp flow, see below),

\item a spatial inhomogeneity of the freeway (such as a ramp,
gradient, or change in the number of usable lanes), and

\item a temporary perturbation of the traffic flow (e.g. due to
lane changes~\cite{MOBIL-TRR07} or long-lasting overtaking maneuvers
of trucks~\cite{Martin-empStates,HelbTilch-EPJB-08}).
\end{enumerate}
The challenge of traffic modeling, however, goes considerably beyond
this. It would be favorable, if the traffic dynamics could be understood on the basis of
elementary traffic patterns \cite{Martin-empStates} such as the ones depicted in Fig.~\ref{fig:elemPatterns}, and if complex traffic patterns (see, e.g., Fig.~\ref{fig:combinedPatterns}) could be understood as combinations of them, considering interaction effects. 

It was proposed that the occurrence of elementary congested
traffic states could be classified and predicted by a phase
diagram~\cite{Phase,Opus}. Furthermore, it was suggested that this phase 
diagram can be derived from the instability diagram of traffic flow and the outflow
from congested traffic. This idea has been
taken up in many other publications, also as a means of studying,
visualizing, and classifying properties of traffic
models~\cite{Lee99,Lee-emp,Barlovic-phasediag,CA-phasediag}. However,
it has been claimed that the phase diagram approach would be
insufficient~\cite{Kerner-book}. While some of the criticism
is due to misunderstandings, as will be shown in Sec.~\ref{sec:reply}, the
classical phase diagrams lack, in fact, the possibility of
``widening synchronized patterns'' (WSP) proposed by Kerner and 
Klenov~\cite{Kerner-Mic}, see Fig. \ref{fig:elemPatterns}(d).

\par
In this paper, we will start in Sec.~\ref{sec:defPhases} with a discussion of the somewhat 
controversial notion of ``traffic phases'' and the clarification that we
use it to distinguish congestion patterns with a qualitatively
different spatiotemporal appearance. 
In Sec. \ref{sec:cong} we will show
that existing models can produce all the empirically observed patterns
of Fig. 1, when simulated in an open system with a bottleneck.
We will then present a derivation and explanation of the idealized, schematic 
phase diagram of traffic states in Sec.~\ref{sec:phasediag}. In
contrast to previous publications,  
 we will assume that the critical density $\rho_{\rm c2}$, at which traffic
becomes linearly unstable, is greater than the density $\rhoQmax$,
where the maximum flow is reached (see Appendix~\ref{app:SourceSink} for details). 
As a consequence, we will find that
``widening synchronized pattern''  {\it do} exist within the phase
diagram approach, even for models with a fundamental
diagram. While this analysis is carried out for single, isolated bottlenecks,  
Sec. \ref{sec:rampsOffOn} will introduce how to generalize it to the case of multi-ramp setups.
In Sec. \ref{sec:phasediagOther}, we will discuss other possible
types of phase diagrams, depending on the stability
properties of the considered model.
Afterwards, in Sec.~\ref{sec:phasediagEmp}, we will present
recent empirical data supporting our theoretical phase diagram. 
Sections~\ref{sec:reply} and~\ref{sec:conclusion} will finally try to
overcome some misunderstandings regarding the phase diagram concept
 and summarize our findings.

\section{\label{sec:defPhases}On the Definition of Traffic Phases}

Before we present the phase diagram of traffic states, it must be
emphasized that some confusion arises from the different use of the
term ``(traffic) phase''. In thermodynamics, a ``phase'' corresponds to an
equilibrium state in a region of the parameter space of thermodynamic
variables (such as pressure and temperature), in which the appropriate
free energy is \textit{analytic}, i.e., all first and higher-order
derivatives with respect to the thermodynamic variables exist. One
speaks of a {\it first-order} phase transition, if a first derivative, or
``order parameter'', is discontinuous, and of a ``second-order'' or
``continuous'' phase transition, if the first derivatives are
continuous but a second derivative (the ``susceptibility'')
diverges. What consequences does this have for defining ``traffic
phases''?
\par
Although traffic flow is a self-driven nonequilibrium system, it has
been shown~\cite{Arndt-nonlinThermo} that much of the equilibrium
concepts can be transferred to driven or self-driven non-equilibrium 
systems by appropriately redefining them. Furthermore, concepts of classical
thermodynamics have been successfully applied to nonequilibrium
physical and nonphysical systems, yielding quantitatively correct
results. This includes, for example, the application of the
fluctuation-dissipation theorem~\cite{FDT} (originally referring to
equilibrium phenomena) to vehicular traffic~\cite{FPE-EuroPhys08}.
\par
In contrast to classical thermodynamics,
{\it nonequilibrium} phase transitions are {\it possible} in one-dimensional
systems~\cite{Eva98}. However, according to the definition of
phase transitions, one needs to make sure that details
of the boundary conditions or finite-size effects do not play a role
for the characteristic properties of the phase. Furthermore, one
must define suitable order parameters or susceptibilities. 
While the first propositions have been already made
a decade ago~\cite{Lubeck-98}, there is no general agreement
regarding the quantity that should be chosen
for the order parameter. Candidates include the density, the fraction of
vehicles in the congested state~\cite{Lubeck-98}, the average velocity or flow,
or the variance of density, velocity, or flow. Whenever one observes a discontinuous or 
hysteretic transition in a large enough system, there is no need to define an order parameter, as 
this already implies a first-order phase transition. For {\it continuous},
symmetry-braking phase transitions, the deviation from the more
symmetric state (e.g. the amplitude of density
variations as compared to the homogeneous state) seems to be
an appropriate order parameter.
\par
To summarize the above points, it appears that thermodynamic phases {\it can},
in fact, be defined for traffic flow. In connection with transitions
between different traffic states at bottlenecks, we particularly
mention the notion of
\textit{boundary-induced phase transitions}~\cite{KrugLetter,schuetz,Popkov-boundaryPhase}.
Here, the boundary conditions have been mainly used as a means to
control the average density in the open system under consideration.
\par
In publications on traffic, a ``phase'' is often interpreted as
``traffic pattern'' or ``traffic state with a typical spatio-temporal
appearance''. Such states depend on the respective boundary conditions. In
this way, models with several 
phases can produce a multitude of spatiotemporal patterns.  It should
become clear from these considerations that the various proposed
``phase diagrams'' do {\it not} relate to {\it thermodynamic} phases, but classify
spatio-temporal states, as is common in systems theory. In these
non-thermodynamic phase diagrams, the ``phase space'' is spanned by certain 
control parameters, e.g. by suitably parameterized boundary conditions, by inhomogeneities
(bottleneck strengths), or by model parameters~\cite{HDM}. 
For example, the phase diagrams discussed in Refs.~\cite{Phase,Kerner-book} 
and this paper contain the
axes ``main inflow'' (i.e., an upstream boundary condition) and
``on-ramp flow'' (characterizing the bottleneck strength).
\par
In any case, {\it empirical} observations of the traffic dynamics 
relate to the spatiotemporal traffic patterns, and not to the
thermodynamic phases. Therefore, the quality of a traffic model should
be assessed by asking whether it can produce all  observed kinds of 
spatio-temporal traffic patterns, including the conditions for their appearance.

\begin{figure*}

\centering
\includegraphics[width=\textwidth]{\figpath{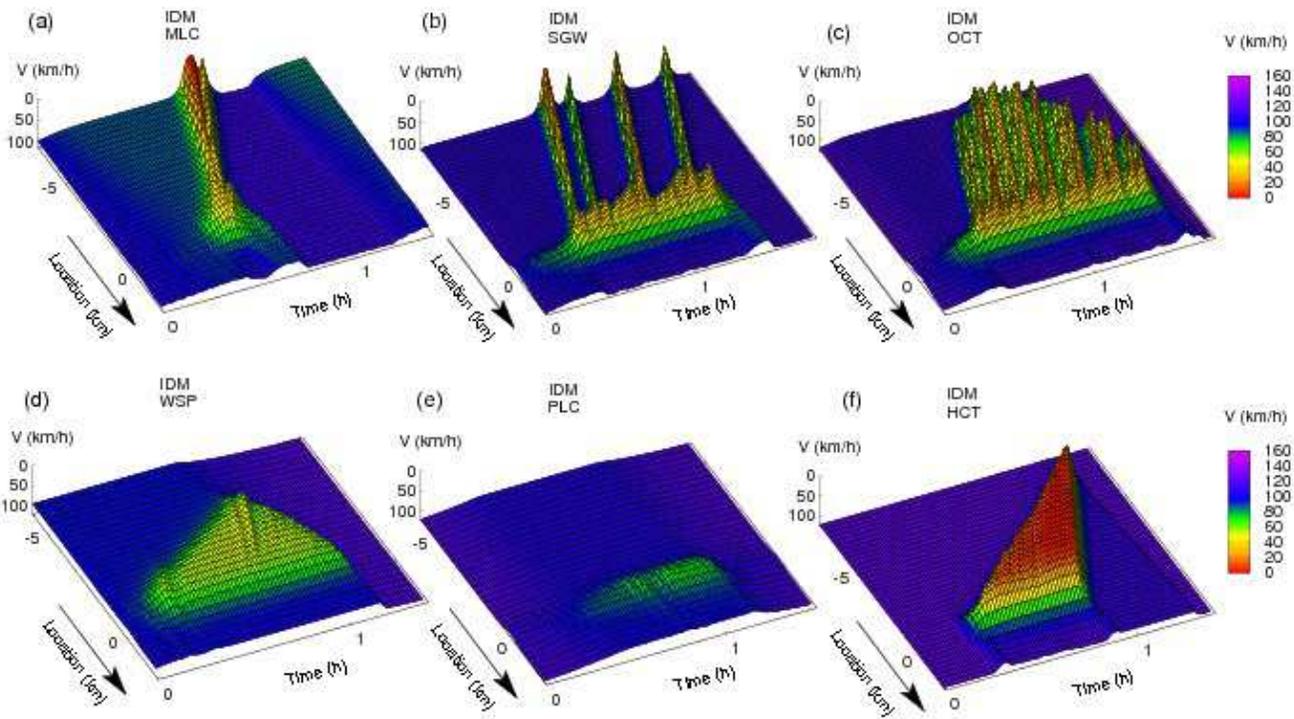}}

 \caption{\label{fig:elemPatternsIDM} Simulation of traffic on a
 freeway with an on-ramp at location $x=0$~km using the intelligent
 driver model (IDM)~\cite{Opus} with parameters corresponding to an
 instability diagram as illustrated in
 Fig.~\protect\ref{fig:stabdiag}(d).  The macroscopic velocity field
 was extracted from the simulated trajectories by placing virtual
 detectors every \unit[500]{m} and determining the velocity with the
 same method~\protect\cite{Treiber-smooth} that has been used for the
 data.  Depending on the respective traffic flows on the ramp and on
 the freeway, different kinds of congested traffic states emerge: a
 moving cluster (MC), a pinned localized cluster (PLC), (``triggered'')
 stop-and-go waves (SGW), oscillating congested traffic (OCT), or
 homogeneous congested traffic (HCT). During the first minutes of the
 simulation, the flows on the freeway and the on-ramp were increased
 from low values to their final values. Since the assumed flows fall
 into a metastable traffic regime, the actual breakdown was initiated
 by additional perturbations of the ramp flow. }
\end{figure*}

\section{\label{sec:cong}Congested Traffic States}

When simulating traffic flow with the ``microscopic'' intelligent
driver model (IDM)~\cite{Opus}, the optimal velocity model
(OVM)~\cite{Bando-98}, the non-local, gas-kinetic-based traffic model
(GKT)~\cite{GKT}, or the ``macroscopic'' Kerner-Konh\"auser
model~\cite{KeKo93} (with the parameter set chosen by Lee {\it et
al.}~\cite{Lee}), we find free traffic flow and different kinds of
congestion patterns, when the ramp flow $Q_{\rm on}$ and the upstream
arrival flow $Q_{\rm up}$ on the freeway are varied. The diversity of
traffic patterns is
\begin{enumerate}
\item due to the possibility of having
either {\it locally constraint} or {\it spatially extended}
congestion\footnote{Note that traffic patterns which appear to be
localized, but continue to grow in size,  belong to the spatially
extended category of traffic states. Therefore, ``widening moving
clusters'' (WMC) are classified as extended congested traffic, while
the similarly looking ``moving localized clusters'' (MLC) are
not. According to Fig. \protect\ref{fig:phasediag2a}, however, the phases of
both states are located next to each other, so one could summarize
both phases by one area representing ``moving clusters'' (MC).}  
and
\item due to the possibility of having stable, unstable or free traffic flows.
\end{enumerate}
Typical representatives of congested traffic patterns obtained by computer simulations
with the intelligent driver model~\cite{Opus} are shown in Fig.~\ref{fig:elemPatternsIDM}. 
Notice that all empirical patterns displayed in Fig. \ref{fig:elemPatterns} can be reproduced.
\par 
One can distinguish the different traffic states (i.e. congestion patterns) by analyzing the temporal {\it and} spatial
dependence of the average velocity $V(x,t)$: If $V(x,t)$ stays above a certain threshold
$V_{\rm crit}$, where $x$ is varied within a homogeneous freeway section
upstream of a bottleneck, we call the traffic state
{\it free traffic} (FT), otherwise congested traffic.\footnote{A typical threshold for German
freeways would be $V_{\rm crit} \approx 80$ km/h.} If these speeds
fall below $V_{\rm crit}$ only over a short freeway subsection, and the
length of this section is approximately stable or stabilizes over
time, we talk about {\it localized clusters} (LC), otherwise of {\it
spatially extended congestion} states  (see also footnote 1).
\par
According to our simulations, there
are two forms of localized clusters: {\it Pinned localized clusters}
(PLC) stay at a fixed location over a longer period of time, while
{\it moving localized clusters} (MLC) propagate upstream with the
characteristic speed $c_0$.  These states have to be contrasted with
extended congested traffic\footnote{which includes ``widening moving
clusters'' (see Fig. \ref{fig:elemPatterns}(a) and footnote 1)}: 
{\it Stop-and-go waves} (SGW) may be
interpreted as a sequence of several moving localized
clusters. Alternatively, they may be viewed as special case of {\it
oscillating congested traffic} (OCT), but with free traffic flows of about $Q_{\rm out}\gtrsim 1800$ vehicles/h/lane between the upstream propagating jams.  Generally,
however, OCT is just characterized by oscillating speeds in
the congested range, i.e. unstable traffic flows. If the speeds are congested over a spatially
extended area, but not oscillating,\footnote{when averaging over
spatial and temporal intervals that sufficiently eliminate effects of
heterogeneity and pedal control in real vehicle traffic} we call this {\it homogeneous
congested traffic} (HCT). It is typically related with low vehicle velocities. 
\par 
In summary, besides free traffic, the above mentioned and some other traffic models
predict five different, spatio-temporal patterns of congested traffic states at a
simple on-ramp bottleneck: PLC, MLC, SGW, OCT, and HCT. Similar
traffic states have been identified for flow-conserving
bottlenecks in car-following models~\cite{Helb-Opus,Helb-micmac}, and
for on-ramps and other types of bottlenecks in macroscopic models
\cite{Phase,Lee}.  

In contrast to this past work, we have also simulated an additional
traffic pattern [Fig.~\ref{fig:elemPatternsIDM}(d)]. This pattern has
a similarity to the  {\it widening synchronized pattern} (WSP)  proposed by 
Kerner in the framework of his three-phase
traffic theory~\cite{BKernerKlenov-02}.
In the following section, we show how this pattern may be understood
in the framework of models with a fundamental diagram.

\section{\label{sec:phasediag}Derivation 
and Explanation of the Phase Diagram of Traffic States}

\begin{figure}
\centering
\includegraphics[width=0.5\textwidth]{\figpath{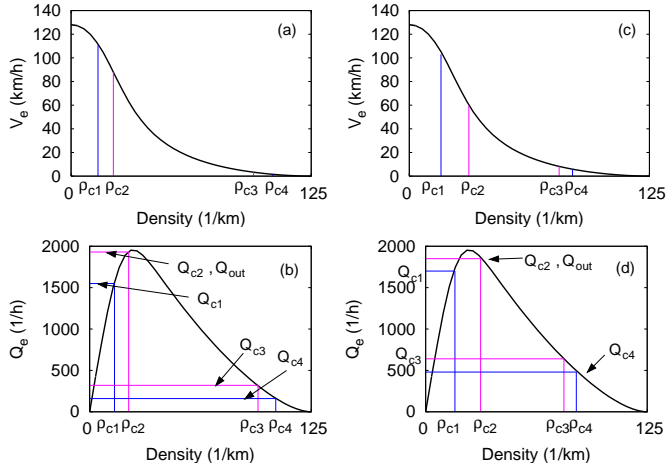}}

  \caption{
  Illustration of stable, linearly
  unstable, and metastable density regimes within velocity-density 
  diagrams $V_{\rm e}(\rho)$ (top) and the flow-density diagrams 
  $Q_{\rm e}(\rho)$ (bottom). Traffic is stable for $\rho<\rho_{c1}$ and
  $\rho>\rho_{c4}$ and linearly unstable for $\rho_{c2}<\rho
  <\rho_{c3}$. These two regimes are separated by a low-density and a
  high-density region of metastable traffic given by the intervals
  $\rho_{c1}<\rho <\rho_{c2}$ and $\rho_{c3}<\rho <\rho_{c4}$,
  respectively. In the metastable regimes, perturbations in the traffic flow grow, if their size
  is larger than a certain critical amplitude~\cite{HelbMoussaid-EPJB-08}, otherwise they fade away. 
  The critical amplitude is largest towards the boundaries $\rho_{\rm c1}$ and 
  $\rho_{\rm c4}$ of unconditionally stable traffic flow, while it goes to zero towards
  the boundaries $\rho_{\rm c2}$ and $\rho_{\rm c3}$ of linearly unstable traffic.
  Note that the metastable and unstable regimes may
  vanish for certain traffic models or parameter specifications. 
  The possible types of congested traffic patterns
  depend on the existence of the different stability regimes and on
  the relative position of their boundaries with respect to the
  density $\rhoQmax$ at capacity $Q_{\rm max}$ (maximum flow). The left figures
  show the situation for $\rho_{c2}<\rhoQmax$, the right figures the situation
  for $\rho_{c2}>\rhoQmax$.}
\label{fig:stabdiag}

\end{figure}

\begin{figure}
\centering
\includegraphics[width=0.24\textwidth]{\figpath{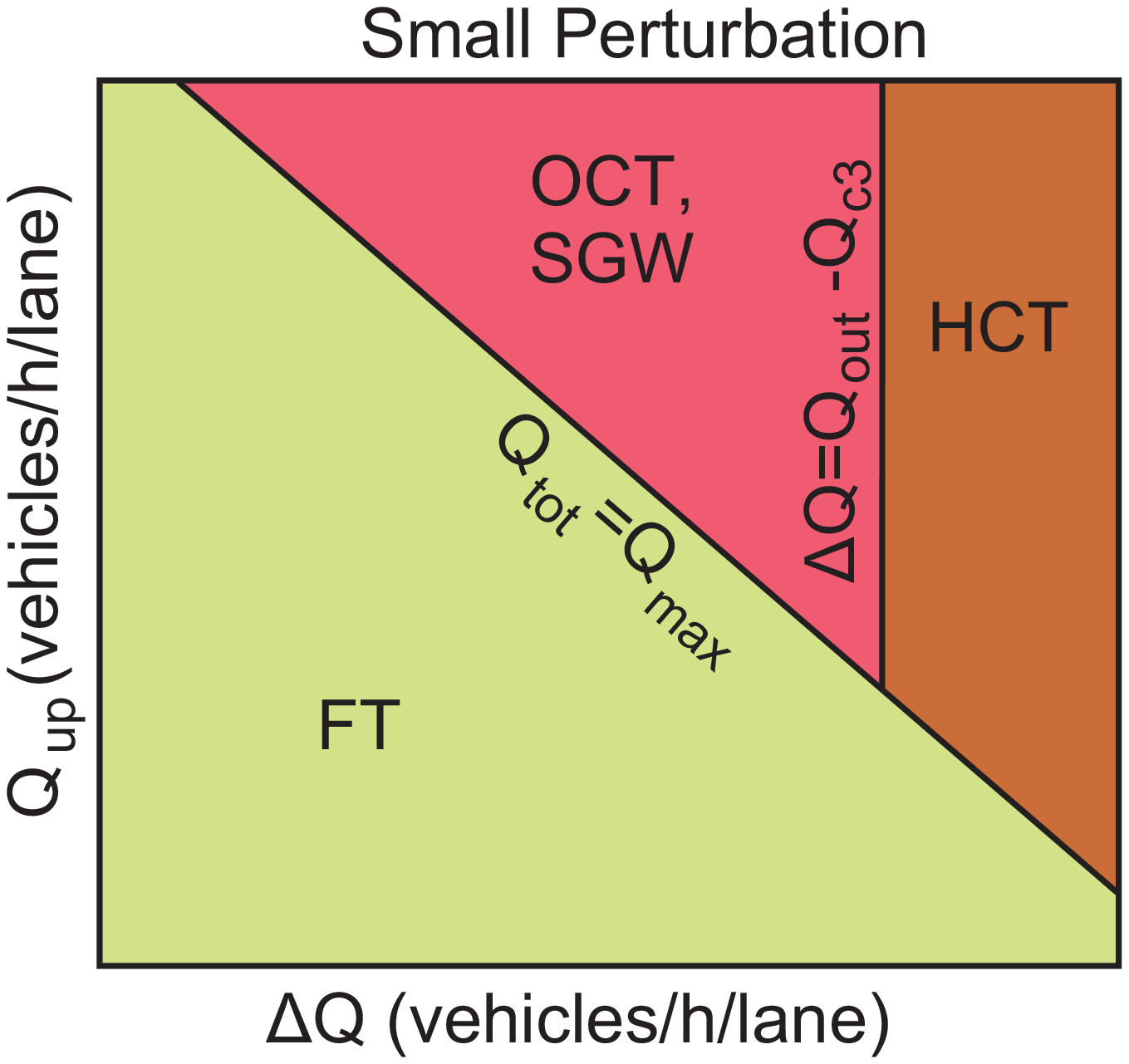}}
\includegraphics[width=0.24\textwidth]{\figpath{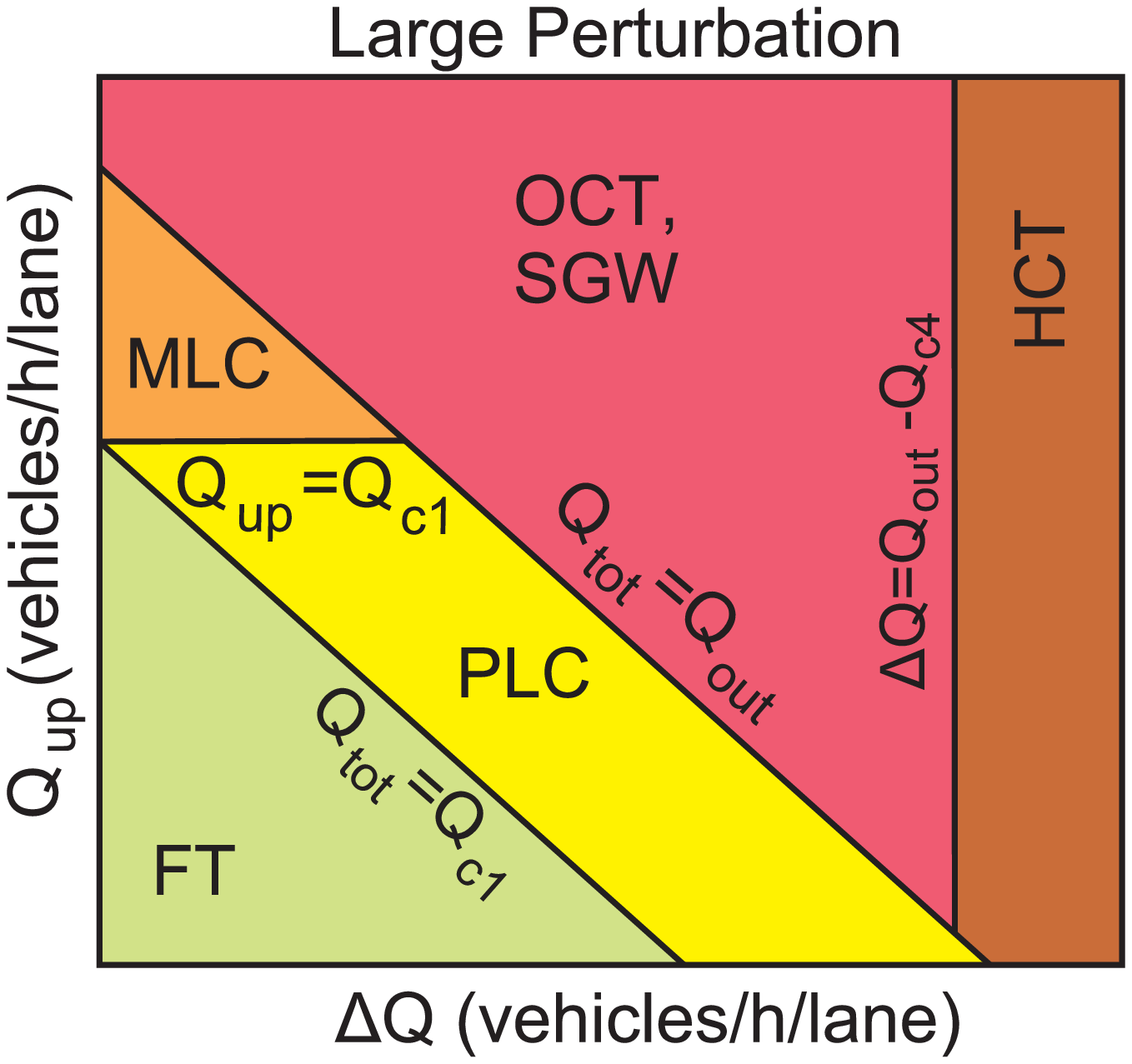}}

  \caption{Schematic (idealized) phase diagrams for the expected traffic patterns 
as a function of the upstream freeway flow $Q_{\rm up}$ and the ramp flow $\Delta Q$,
as studied in Refs.~\protect\cite{Phase,Opus}. The left figure is for negligible, the right
figure for large perturbations. The situation for medium-sized
perturbations can lie anywhere between these two extremes. For
example, in the area marked as PLC, one may find free traffic or
pinned localized clusters, or in some of the area attributed to HCT,
one may find SGW or OCT states. The assumed instability diagram
underlying the above schematic phase diagrams is depicted in
Figs.~\ref{fig:stabdiag}(a) and (b). With $\rho_{\rm c1} < \rho_{\rm c2} <
\rhoQmax<\rho_{\rm c3} < \rho_{\rm c4} < \rho_\text{jam}$, it assumes
no degeneration of the critical densities $\rho_{{\rm c}k}$ and a
stable flow at high densities. Note that, for illustrative reasons, we
have set aside the exact correspondence of the flow values $Q_{{\rm
c}k}$.}

\label{fig:phasediag1a}
\end{figure}

\begin{figure}
\centering
\includegraphics[width=0.24\textwidth]{\figpath{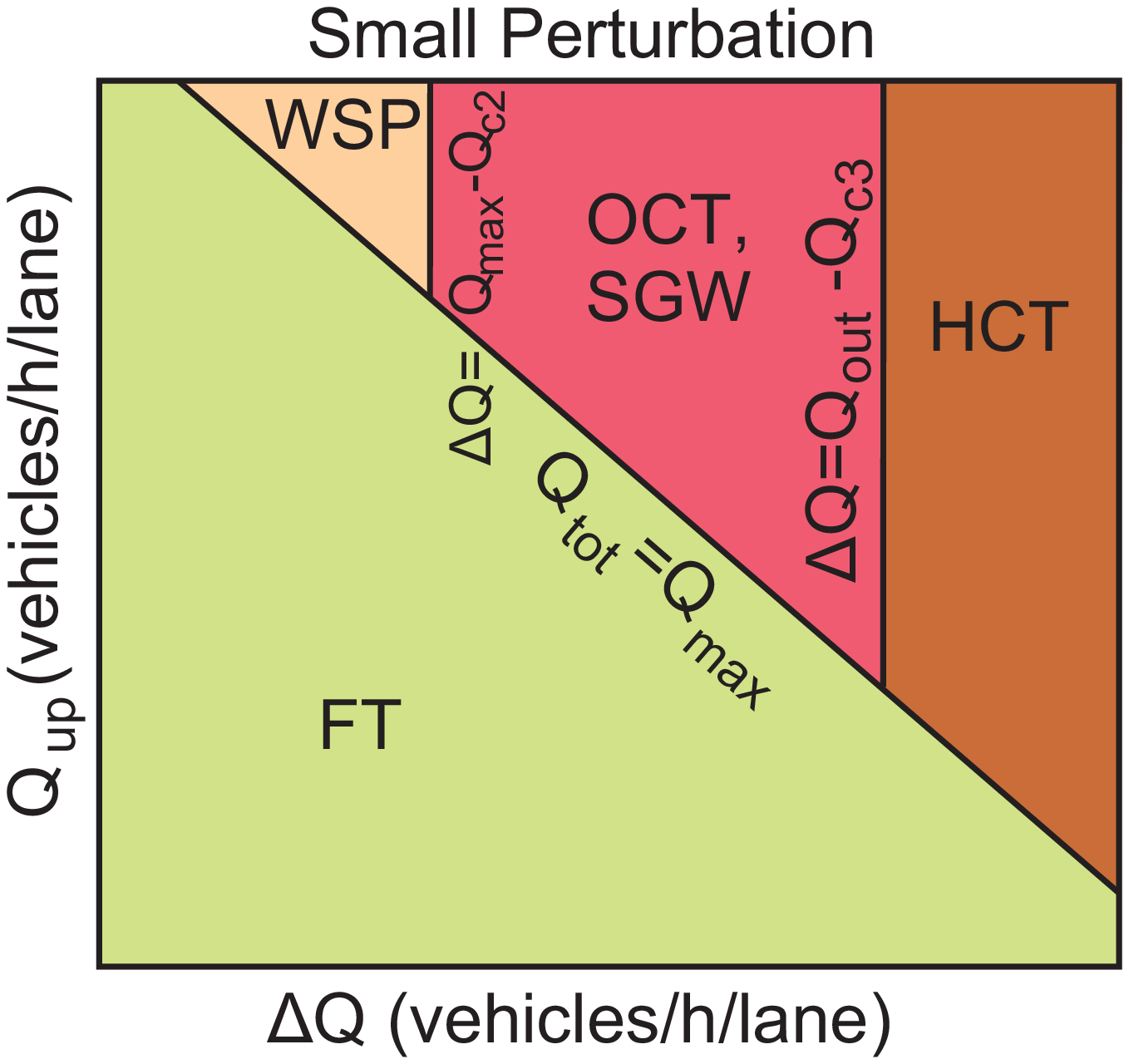}}
\includegraphics[width=0.24\textwidth]{\figpath{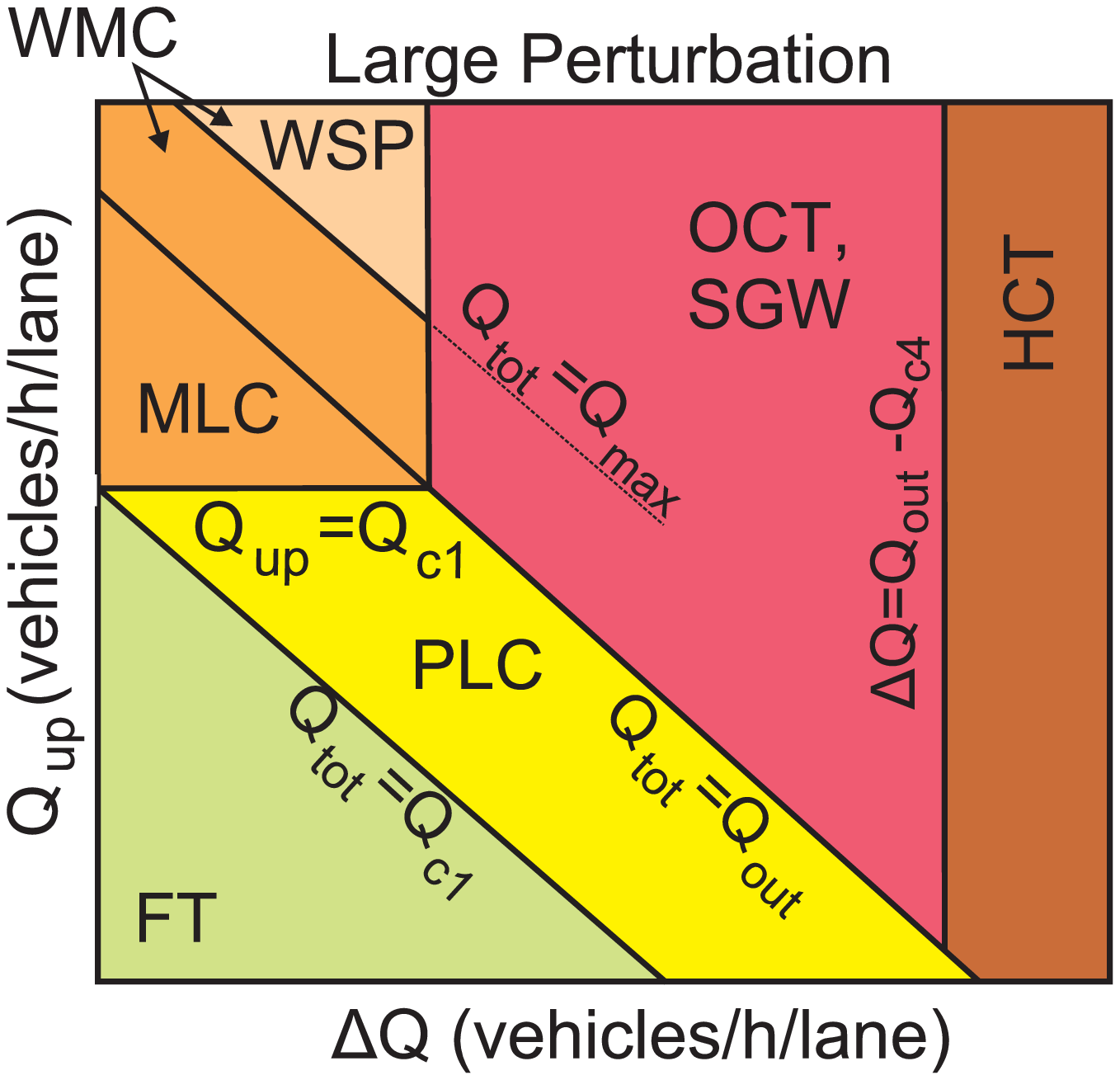}}

\caption{Schematic phase diagram as in Fig.~\ref{fig:phasediag1a}, but for 
the instability diagram represented by Figs.~\ref{fig:stabdiag}(c) and (d).
In contrast to Fig.~\ref{fig:phasediag1a}, traffic flow at capacity is metastable
($\rho_{c1}<\rhoQmax<\rho_{c2}$), which leads to a greater
variety of traffic states in the upper left corner of the phase
diagram. In particularly, we find ``widening synchronized patterns'' (WSP). 
``OCT, SGW'' means that one expects to find oscillating
congested traffic or stop-and-go waves, but not necessarily
both. Together with ``widening moving clusters'' (WMC, see footnote 1) they form the
area of extended oscillatory congestion. However, the WMC and MLC
phases may also be summarized by one area representing ``moving
clusters'' (MC).}
\label{fig:phasediag2a}
\end{figure}

It turns out that the possible traffic patterns in open systems with
bottlenecks are mainly determined
by the instability diagram (see Fig. \ref{fig:stabdiag}), no matter if the model
is macroscopic or microscopic.  This seems to apply at least for traffic models with a fundamental diagram, which we will focus on in the following sections. Due to the close relationship with the instability diagram, the preconditions for the
possible occurrence of the different traffic states can be illustrated
by a {\it phase diagram}. Figures~\ref{fig:phasediag1a} and \ref{fig:phasediag2a} show two examples. Each area of a phase diagram represents the combinations of upstream freeway
flows $Q_{\rm up}$ and bottleneck strengths $\Delta Q$,
for which a certain kind of traffic state can exist. 
\par
It is obvious that an on-ramp flow $Q_{\rm on}(t)$, for example,  
causes a bottleneck, as it consumes some of the capacity of the freeway. 
$Q_{\rm on}(t)$ represents the flow actually entering the freeway via
the on-ramp, i.e. the flow {\it leaving} the on-ramp and not the flow
{\it entering} the on-ramp.\footnote{When the freeway is busy, it
may happen that these two flows are different and that a queue of
vehicles forms on the on-ramp. Of course, it is an interesting
question to determine how the entering ramp flow depends on the
freeway flow, but this is not the focus of attention here, as this
formula is not required for the following considerations.}  We
assume that $Q_{\rm on}(t)$ is known  through a suitable measurement.
Having clarified this, we define the {\it bottleneck strength} due to
an on-ramp by the entering ramp flow, divided by the number $I_{\rm
fr}$ of freeway lanes:
\begin{equation}\label{eq:bottleneck_strength}
  \Delta Q(t) = \Delta Q_{\rm on}(t) = \frac{Q_{\rm on}(t)}{I_{\rm fr}} \, . 
\end{equation}
This is done so, because the average flow $\Delta Q$ added to
each freeway lane by the on-ramp flow corresponds to the capacity that
is not available anymore for the traffic flow $Q_{\rm up}$ coming from
the upstream freeway section.  As a consequence, congestion may form
upstream of the ramp. In the following, we will have to determine the
density inside the forming congestion pattern and where in the
instability diagram it is located. It will turn out that, given
certain values of $Q_{\rm up}$ and $\Delta Q$, the different regions
of the phase diagram can be related with the respectively observed or
simulated spatiotemporal patterns.  We distinguish free traffic and
different kinds of localized congested traffic as well as different
kinds of extended congested traffic. When contrasting our classification
of traffic states with Kerner's one \cite{Kerner-book}, we find the
following comparison helpful:
\begin{enumerate}
\item According to our understanding, what we call 
``extended congested traffic'' 
 may be associated with Kerner's ``synchronized flow''.
In particular, the area where Kerner's phase diagrams predict a
``general pattern'' 
matches well with the area, where we expect OCT and HCT states.
\item  ``Moving clusters''\footnote{i.e. ``moving localized clusters'' (MLC) and ``widening moving clusters''
(WMC), see footnote 1 and Sec. \ref{sec:classicalPatterns}} may be
associated with ``wide moving jams'' and/or ``moving synchronized
patterns'' (MSP). 
\item ``Stop-and-go waves'' appear to be related with
multiple ``wide moving jams'' generated by the ``pinch effect''.
\item   ``Pinned localized clusters'' may related to  
Kerner's ``localized synchronized pattern'' (LSP).
\item  Kerner's ``widening synchronized pattern'' (WSP) and ``dissolving general pattern'' (DGP)
did not have a correspondence with results of our own computer
simulations so far.  
These states are predicted to appear for high freeway flows and low
bottleneck strengths.  
In the following subsections, we report that, quite unexpectedly,
similar results are found for certain traffic models having 
a fundamental diagram. 
\end{enumerate}
The phase diagram
can not only be determined {\it numerically}.  It turns out that the
borderlines between different areas (the so-called phase boundaries)
can also be {\it theoretically} understood, based on the flows
\begin{equation}
 Q_{{\rm c}k} = Q_{\rm e}(\rho_{{\rm c}k})
\end{equation}
at the instability thresholds $\rho_{{\rm c}k}$ ($k=1, \ldots, 4$), the maximum flow
capacity $Q_{\rm max}$ under free flow conditions, and the dynamic
flow capacity, i.e. the characteristic outflow ${Q}_{\rm out}$ from
congested traffic~\cite{Kerner-Rehb96} (see Fig.~\ref{fig:stabdiag}). 
$Q_{\rm e}(\rho)$ represents the equilibrium flow-density relationship, 
which is also called the ``fundamental diagram''.
\par

The exact shape and location of the separation lines
between different kinds of traffic states depend on the traffic model and
its parameter values.\footnote{Since the model parameters characterize
the prevailing driving style as well as external conditions such as weather
 conditions and speed limits, the separation
lines (``phase boundaries'') and even the existence of certain traffic
patterns are subject to these factors, see
Sec. \protect\ref{sec:phasediagEmp}.
}
Furthermore, the
characteristic outflow ${Q}_{\rm out}$ typically depends on the type
and strength of the bottleneck.\footnote{For example, in most models, 
the outflow $Q_{\rm out}$ downstream of an on-ramp bottleneck
decreases with the bottleneck strength and increases with the length of the
on-ramp~\protect\cite{Opus,Treiber-ThreePhasesTRB}.}
For the sake of simplicity of our discussion, however, we will assume constant
outflows $Q\sub{out}$ in the following.
\par
The meaning of the different critical density thresholds $\rho_{{\rm
c}k}$ and flow thresholds $Q_{{\rm c}k} = Q_{\rm e}(\rho_{{\rm c}k})$,
respectively, is described in the caption of
Fig. \ref{fig:stabdiag}. Note that the density $\rho_{{\rm c}2}$ may
be smaller or larger than the density $\rhoQmax$ at capacity, where
the maximum flow $Q_{\rm max}$ is reached. Previous computer
simulations and phase diagrams mostly assumed parameters where traffic
at capacity is linearly unstable ($\rho_{{\rm c}2} <
\rhoQmax<\rho_{{\rm c}3}$), which is depicted in
Figs. \ref{fig:stabdiag}(a) and (b).  However, in some traffic models
such as the IDM \cite{Opus}, the stability thresholds can
be controlled in a flexible way by varying their model parameters (see
Appendix~\ref{app:IDMstab}). In the following, we will focus on the
case where traffic at capacity is metastable ($\rho\sub{c2}> \rhoQmax
>\rho\sub{c1}$), cf. Fig. \ref{fig:stabdiag}(c) and (d).\footnote{The
IDM parameters for plots (a) and (b) are given by
$v_0=\unit[128]{km/h}$, $T=\unit[1]{s}$, $s_0=\unit[2]{m}$,
$s_1=\unit[10]{m}$, $a=\unit[0.8]{m/s^2}$, and
$b=\unit[1.3]{m/s^2}$. To generate plots (c) and (d), the acceleration
parameter was increased to $a=\unit[1.3]{m/s^2}$, while the other
parameters were left unchanged.  } As will be shown in the next
subsection, this appears to offer an alternative explanation of the
``widening synchronized pattern'' (WSP) introduced in
Ref.~\cite{BKernerKlenov-02}, see Fig.~\ref{fig:elemPatternsIDM}(d).
Simpler cases will be addressed in Sec.~\ref{sec:phasediagOther}
below.

\subsection{\label{sec:WSP}Transition to Congested Traffic for Small Bottlenecks}

In the following, we restrict our considerations to situations with
one bottleneck only, namely a single on-ramp. Combinations of off- and
on-ramps are not covered by this section. They will be treated later
on (see Sec.~\ref{sec:rampsOffOn}).
\par
For matters of illustration, we assume a typical rush hour scenario, in which the total traffic volume
\begin{equation}
 Q_{\rm tot}(t) = Q_{\rm up}(t) + \Delta Q(t) \, ,
\end{equation}
i.e. the sum of the flow $Q_{\rm up}(t)$ sufficiently upstream of the
ramp bottleneck and the on-ramp flow $\Delta Q(t)$ per freeway lane, is
increasing with time $t$. As long as traffic flows freely, the flow downstream of
the bottleneck corresponds to the total flow $Q_{\rm tot}(t)$, while
the upstream flow is $Q_{\rm up}(t)$. 
\par
When the total flow $Q_{\rm tot}(t)$ exceeds the critical density $\rho_{\rm c1}$, it enters the metastable density regime. That is,
large enough perturbations may potentially grow and cause a breakdown
of the traffic flow. However, often the perturbations remain
comparatively small, and the total traffic volume rises
so quickly that it eventually exceeds the maximum freeway capacity 
\begin{equation}
\label{cond-dissolvWSP}
Q_{\rm tot} = Q_{\rm up} + \Delta Q > Q_{\rm max} = \max_\rho  Q_{\rm e}(\rho) 
= Q_{\rm e}(\rho_{\rm max}) \, . 
\end{equation}
This is reflected in the left phase diagram in Fig. \ref{fig:phasediag2a}
by the diagonal line separating the states ``FT'' and ``WSP''. (Note that $\rho_{\rm max}$ represents the density, for which the maximum free traffic flow occurs, not the jam density $\rho_{\rm jam}$.) 
\par
When the total traffic volume $Q_{\rm tot}$ exceeds the maximum capacity $Q_{\rm max}$, 
a platoon of vehicles will form upstream of the
bottleneck. Since, in this section, we assume {\it metastable} traffic at maximum capacity $Q_{\rm max}$ 
 (see Fig. 7 in Ref.~\cite{HelbTilch-EPJB-08}), this will not instantaneously  lead to a
traffic breakdown with an associated capacity drop. Thus, the  flow downstream 
of the bottleneck remains limited to
$Q_{\rm max}$ (at least temporarily). As the on-ramp flow takes away an amount $\Delta Q$
of the maximum capacity $Q_{\rm max}$, the (maximum) flow upstream of the
bottleneck is given by  
\begin{equation}
\label{Qbot}
 Q_{\rm bot} = Q_{\rm max} - \Delta Q \, .
\end{equation}
When the actual upstream flow $Q_{\rm up}$ exceeds this value, a mild form of congestion will result upstream of the ramp.
The density of the forming vehicle platoon is predicted to be 
\begin{equation}
\label{rhobot}
\rho\sub{bot}=\rho_{\rm cg}(Q_{\rm bot}) = \rho_{\rm cg}\big(Q_{\rm
max} - \Delta Q\big) >  \rho_{\rm max}\, , 
\end{equation}
where $\rho_{\rm cg}(Q)$ is the density corresponding to a stationary and homogeneous
{\it congested} flow of value $Q$ (i.e. it is the inverse function of the ``congested branch''
of the fundamental diagram).
\par
According to the equation for the propagation speed of shockwaves (see
Ref.~\cite{Whitham74}),  the upstream front of the
forming vehicle platoon is expected to propagate upstream at the speed
\begin{equation}
 C_1(t) = \frac{Q_{\rm up} - Q_{\rm bot}}
{\rho_{\rm fr}(Q_{\rm up}) - \rho_{\rm cg}(Q_{\rm bot})} \, ,
\label{C1}
\end{equation}
where $\rho_{\rm fr}(Q)$ is the density of stationary  and homogeneous
traffic at a given flow $Q$ (i.e. the inverse function of the ``free branch'' of the
fundamental diagram.)

\begin{figure*}
\centering
\includegraphics[width=\textwidth]{\figpath{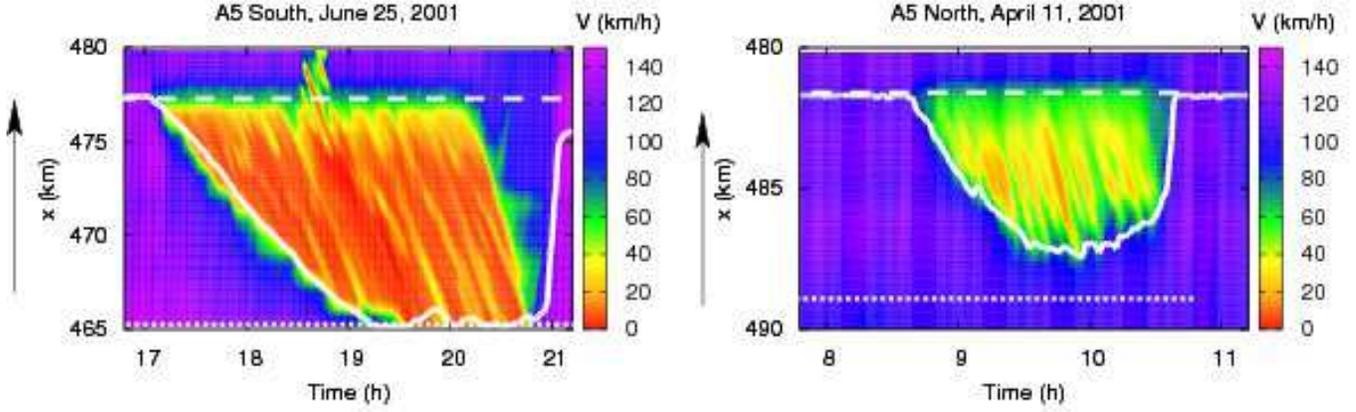}}

  \caption{\label{fig:upFront}Examples of congestion patterns on the
  German freeway A5 close to Frankfurt, for which data have been
  provided to the authors between kilometers 465 and 492. The figures
  analyze the propagation of the upstream front of a region of
  congested traffic (white solid line) according to
  Eq.~\protect\eqref{C2}, for empirical HCT (left) and OCT (right).
  In both plots, the driving direction is upwards (as indicated by the
  arrows).  The upstream flow $Q_\text{up}$ was determined from a
  detector cross section whose location is indicated by a dotted white
  line, while the bottleneck flow was determined from detectors of a
  nearby cross section (dashed white line). When determining the
  flows, the time delay caused by the finite propagation velocities
  $dQ_\text{e}(\rho)/d\rho$ from the detectors to the upstream front
  was taken care of.  The congestion patterns were chosen such that
  there were no ramps at or between the two detector cross
  sections. Otherwise, the determination of $Q_\text{up}$ and
  $Q_\text{cong}$ would have been more complicated. The free and
  congested densities were calculated with a simple, triangular
  fundamental diagram. Therefore, $\rho_\text{fr}(Q)=Q/V_0$ and
  $\rho_\text{cg}(Q)=\rho_\text{jam}(1-QT)$, where the following
  parameters were chosen: $V_0=\unit[120]{km/h}$,
  $\rho_\text{jam}=\unit[100]{veh/km/lane}$, and $T=\unit[2]{s}$. }

\end{figure*}

Note that this high-flow situation can persist for a significant time period
only, if the flow $Q_{\rm bot}$ in the platoon is stable or
metastable. This is the case if one of the following applies:
\begin{itemize}
\item[(i)] The traffic flow is unconditionally stable for all
densities such as in the Lighthill-Whitham model \cite{Lighthill-W}. This will
be discussed in Sec. \ref{sec:phasediagOther} below.
\item[(ii)] Traffic flow at capacity is metastable and the bottleneck
is sufficiently weak. This gives rise to the {\it widening synchronized
pattern} (WSP), as will be discussed in the rest of this subsection.
\end{itemize}
By WSP, we mean a semi-congested extending traffic state without large-scale
oscillations or significant velocity drops below, say,
\unit[30-40]{km/h} \cite{Kerner-book}. Putting aside stochastic
accelerations or heterogeneous driver-vehicle populations, this
corresponds to (meta-)stable vehicle platoons at densities
greater than, but close to the density $\rhoQmax$ at capacity. 
This can occur when $\rho_{\rm bot}$ lies in the metastable density
range, i.e. $\rhoQmax <\rho\sub{bot}<\rho_{\rm c2}$,  corresponding to
$Q_{\rm max} > Q_{\rm bot} = Q_{\rm max} - \Delta Q > Q_{\rm c2}$ or
\begin{equation}
 \Delta Q < Q_{\rm max} - Q_{\rm c2} . \label{thiscond}
\end{equation}
In Fig.~\ref{fig:phasediag2a}, this condition belongs to the area left of the
vertical line separating the WSP and OCT states. 
If the bottleneck strength $\Delta Q$ becomes
greater than $Q_{\rm max} - Q_{\rm c2}$, or if $\rho_{\rm cg}(Q_{\rm
bot})$ lies in the metastable regime and perturbations in the
traffic flow are large enough, traffic flow becomes unstable and
breaks down. After the related capacity drop by the amount
\begin{equation}
\Delta Q_{\rm drop} = Q_{\rm max} - Q_{\rm out} \, , 
\end{equation}
the new, ``dynamic'' capacity will be given by the outflow $Q_{\rm out}$ 
from congested traffic \cite{Opus}. 
Obviously, the capacity drop causes the formation of more serious
congestion\footnote{and the condition $Q\sub{up}+\Delta Q<Q\sub{out}$ 
for the gradual {\it dissolution} of the resulting congestion pattern is harder to fulfil than the 
condition $Q_{\rm up} + \Delta Q < Q_{\rm max}$ implied by
Eq.~\eqref{cond-dissolvWSP}}. 
This is illustrated in the right phase diagram of Fig. \ref{fig:phasediag2a} by the offset
between the diagonal lines separating free traffic from WSP and
the other extended congested states (OCT, SGW, and HCT). 
In the following, we will focus on the traffic states {\it
after} the breakdown of freeway capacity from $Q_{\rm max}$ to $Q_{\rm
out}$ has taken place.

\subsection{\label{sec:classicalPatterns}Conditions for Different Kinds of
Congested Traffic after the Breakdown of Traffic Flow} 

For the sake of simplicity, we will assume the case
\begin{equation}
 Q_{\rm c4} < Q_{\rm c3} < Q_{\rm c1} \le Q_{\rm out} \le Q_{\rm c2} <
Q_{\rm max} \, , 
\end{equation}
which seems to be appropriate for real traffic (particularly in
Germany). However, depending on the choice of model parameters, other
cases are possible. The conclusions may be different, then, but the
line of argumentation is the same. In the following, we will
again assume $\rho_{\rm c2} \ge \rhoQmax$, so that the maximum
flow $Q_{\rm max}$ is metastable. Therefore, it can persist
for some time, until the maximum flow state is destabilized by
perturbations or too high traffic volumes $Q_{\rm tot}(t)$,
which eventually cause a breakdown of the traffic flow. (For $\rho_{\rm
c2} < \rhoQmax$, the capacity drop happens automatically, 
whenever $Q_{\rm tot}(t) > Q_{\rm max}$.)
\par
After the breakdown of traffic flow, the traffic situation downstream is given by the outflow $Q_{\rm out}$ from (seriously) congested traffic. As the actually entering ramp flow requires the capacity $\Delta Q$ per lane, the flow upstream of the bottleneck is limited to
\begin{equation}
Q_{\rm cong} = Q_{\rm out} - \Delta Q \, .
\label{Qcong}
\end{equation}
In analogy to Eq.~\eqref{C1}, the upstream front of this congested flow is
expected to propagate with the velocity 
\begin{equation}
 C_2(t) = \frac{Q_{\rm up} - Q_{\rm cong}}{\rho_{\rm fr}(Q_{\rm up}) -
\rho_{\rm cg}(Q_{\rm cong})} \, , 
 \label{C2}
\end{equation}
 as the upstream freeway flow $Q_{\rm up}$ is assumed to be free. The
downstream end of the congested flow $Q_{\rm cong}$ remains located at the
bottleneck~\cite{helbing-sectionbased03}. 

Figure~\ref{fig:upFront} shows that the propagation of the upstream
front according to Eq.~\eqref{C2} agrees remarkably well with
empirical observations, not only for homogeneous congested flow but
also for the OCT pattern. Since the location of the congestion front
is given by {\it integration} of Eq.~\eqref{C2} over time,
oscillations of the input quantities of this equation are
automatically averaged out.

The resulting congestion pattern depends on the stability properties
of the vehicle density  
\begin{equation}
\rho_{\rm cong}=\rho_{\rm cg}(Q_{\rm cong}) = \rho_{\rm cg} \big(Q_{\rm out} - \Delta Q\big)
\end{equation}
in the congested area,  where the outflow $Q_{\rm out}$ from seriously congested traffic represents the effective freeway capacity under congested conditions and $\Delta Q$ the capacity taken away by the bottleneck. In view of this stability dependence, let us now discuss the meaning of the critical densities $\rho_{{\rm c}k}$ or associated flows $Q_{{\rm c}k} = Q_{\rm e}(\rho_{{\rm c}k})$, respectively, for the phase diagram. 
\par
If $\rho_{\rm c2} < \rho_{\rm cong} < \rho_{\rm c3}$, we expect
unstable, oscillatory traffic flow (OCT or SGW). For $\rho_{\rm c3}
\le \rho_{\rm cong} < \rho_{\rm c4}$, the congested flow is
metastable, i.e. it depends on the perturbation amplitude: One may
either have oscillatory patterns (for large enough perturbations) or
homogeneous ones (for small perturbations). Moreover, 
for $\rho_{\rm cong} \ge \rho_{\rm c4}$ (given that the critical
density $\rho_{\rm c4}$ is smaller than $\rho\sub{jam}$), 
we expect homogeneous, i.e. non-oscillatory traffic flows. 
\par
Expressing this in terms of flows rather than densities, 
one would expect the following: Oscillatory congestion patterns (OCT
or SGW) should be possible for $Q_{\rm c2} > Q_{\rm cong} = Q_{\rm out} - \Delta Q
> Q_{\rm c4}$, i.e. in the range 
\begin{equation}
 Q_{\rm out} - Q_{\rm c2}  <  \Delta Q < Q_{\rm out} - Q_{\rm c4} \, ,\label{stern}
\end{equation}
where we have considered $Q_{\rm c2} \ge Q_{\rm out}$. 
\par
 The assumption that the densities between $\rho_{\rm out}$ with
$Q_{\rm e}(\rho_{\rm out}) = Q_{\rm out}$ and $Q_{\rm max}$ are {\it
metastable}, as we assume here, has interesting implications: A linear
{\it instability} would cause a single moving cluster to trigger
further local clusters and, thereby, so-called ``triggered stop-and-go
waves'' (TSG or SGW) \cite{Phase}. A metastability, in contrast, can suppress
the triggering of additional moving clusters, which allows the
persistence of a {\it single} moving cluster, if the bottleneck
strength $\Delta Q$ is small. As, for $Q_{\rm tot} >  Q_{\rm out}$,
the related flow conditions fall into the area of extended congested
traffic, the spatial extension of such a cluster will grow. Therefore,
one may use the term ``widening moving cluster'' (WMC). 
\par
Furthermore, according to our computer simulations, the capacity
downstream of a widening moving cluster may eventually revert  from
$Q\sub{out}$ to $Q\sub{max}$. This happens in the area, where
``widening synchronized patterns'' (WSP) can appear.\footnote{In this
connection, it is interesting to remember Kerner's ``dissolving
general pattern'' (DGP), which is predicted under similar flow
conditions.} 
Therefore, rather than by Eq. (\ref{stern}), the bottleneck strengths characterizing
OCT or SGW states are actually given by
\begin{equation}
Q_{\rm max} - Q_{\rm c2} <  \Delta Q < Q_{\rm out} - Q_{\rm c4} \, ,
\end{equation}
where the lower boundary corresponds to the boundary of the WSP state, see Eq. (\ref{thiscond}). We point out that a capacity reversion despite congestion
is a special feature of traffic models with $\rho\sub{c2}>\rhoQmax$.
\par
\textit{Homogeneous} congested traffic (the definition of which does not 
cover the homogeneous WSP state) is expected to be possible for 
$Q_{\rm cong} = Q_{\rm out} - \Delta Q < Q_{\rm c3}$, i.e. (meta-) stable flows at high
densities. This corresponds to
\begin{equation}
  \Delta Q > Q_{\rm out} - Q_{\rm c3} \, .
\end{equation}
The occurrence of {\it extended} congested traffic like HCT and OCT
requires an additional condition: The total flow  
must exceed the freeway capacity $Q_{\rm out}$ during serious congestion,\footnote{One may also
analyze the situation with the shock wave equation: Spatially expanding congested traffic results, if the speed of the downstream shock front of the congested area (which is usually zero) minus the speed of the upstream shock front (which is usually negative) gives a positive value.}  
i.e. we must have
\begin{equation}
 Q\sub{tot}=Q_{\rm up} + \Delta Q> Q_{\rm out} \, . \label{twent}
\end{equation}
{\it Localized} congestion patterns,  in contrast, require $Q_{\rm tot} \le Q_{\rm out}$ and 
can be triggered for $Q_{\rm tot}> Q_{\rm c1}$, which implies
\begin{equation}
Q_{\rm c1} < Q_{\rm tot}=Q_{\rm up} + \Delta Q  \le Q_{\rm out} \, . 
\end{equation} 
We can distinguish at least two cases:   On the one hand, if 
\begin{equation}
\label{QupMLC}
 Q_{\rm c1} < Q_{\rm up} < Q\sub{max}\, ,
\end{equation}
the flow upstream of the congested area is {\it metastable,} which allows
jams (and large enough perturbations) to propagate upstream. In this
case, we speak of
{\em moving localized clusters} (MLC). Their propagation speed 
$c_0 =-15\pm 5$ km/h is given by the slope of the jam
line~\cite{Kerner-98e}. 
\par
On the other hand, if 
\begin{equation}
\label{QupNoMLC}
 Q_{\rm up} \le Q_{\rm c1} 
\end{equation}
or $\rho_{\rm fr}(Q_{\rm up}) < \rho_{\rm c1}$, traffic flow upstream of the bottleneck
is {\it stable}. Under such conditions, perturbations and, in particular, localized congestion patterns cannot propagate 
upstream, and they stay at the location of the bottleneck. In this case, one speaks of
{\it pinned localized clusters} (PLC).\footnote{ Since pinned
localized clusters rarely constitute a {\it maximum} perturbation,
they can also occur at higher densities and flows, as long as  $Q_{\rm
up} < Q_{\rm c2}$. Therefore,   
MLC and PLC states can {\it coexist} in the range 
$Q_{\rm c1} < Q_{\rm up} < Q_{\rm c2}$.
For most traffic models and bottleneck types, congestion patterns with 
$Q_{\rm tot} \approx Q_{\rm c1}$ do {\it not} exist, since
{\it localized} congestion patterns do not correspond to {\it maximum}
perturbations. The actual lower boundary $\tilde{Q}_{\rm c1}$
for the overall traffic volume $Q\sub{tot}$  that generates congestion 
is somewhat higher than $Q_{\rm c1}$, but usually lower than $Q_{\rm c2}$.
Considering the metastability of traffic flow in this range and the decay of the
critical perturbation amplitude from $\rho_{\rm c1}$ to $\rho_{\rm c2}$~\cite{HelbMoussaid-EPJB-08},
this behavior is expected. However, for some models and
parameters, one may even have $\tilde{Q}\sub{c1}>Q\sub{out}$. In such cases,
PLC states would not be possible under {\it any} circumstances.}

We underline that the actual outflow $\tilde{Q}_{\rm out}$  from
localized clusters 
corresponds, of course, to their inflow $Q_{\rm up} + \Delta Q$ (otherwise they would 
grow or shrink in space). Therefore, the actual outflow $\tilde{Q}_{\rm out}$
of localized congestion patterns can be smaller than $Q_{\rm out}$, i.e. smaller than
the outflow of {\it serious} congestion.  

\section{\label{sec:rampsOffOn}Combinations of On- and Off-Ramps}

\begin{figure}[t!]
\centering
\includegraphics[width=0.5\textwidth]{\figpath{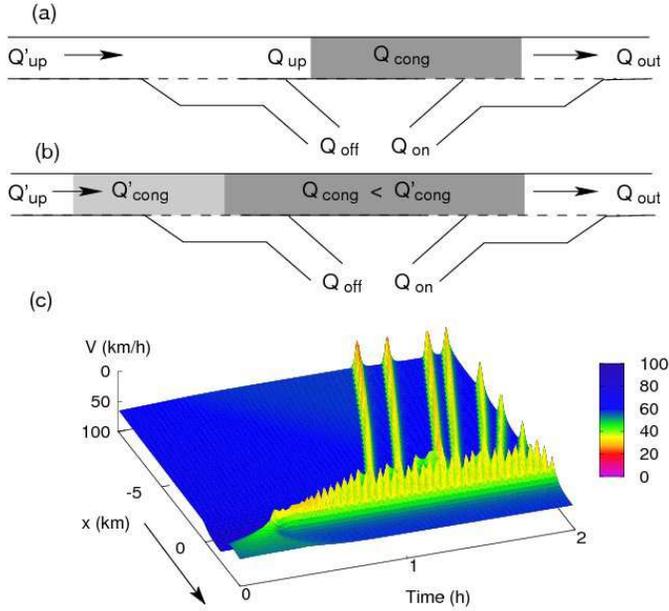}}
  
  \caption{Combination of an on-ramp bottleneck with an upstream
  off-ramp.   (a) When the flow $Q_{\rm up} = Q'_{\rm up} - \Delta Q_{\rm off}$ upstream of the on-ramp
  exceeds $Q_{\rm cong}$, which is defined as the outflow $Q_{\rm out}$ from congested traffic 
  minus the on-ramp flow $\Delta Q_{\rm on} = Q_{\rm on}/I_{\rm fr}$, 
  congested traffic upstream of the on-ramp (dark grey area) is
expected to grow. (b) As soon as the congested area extends up to the
location of the off-ramp, the off-ramp bottleneck is activated. Its effective
outflow $Q'_{\rm out}$ is given by the congested flow $Q_{\rm cong}$
upstream of the on-ramp, while congested flow $Q'_{\rm cong}$
upstream of the off-ramp is higher by the amount $\Delta Q_{\rm off} = Q_{\rm off}/I_{\rm fr}$
of the off-ramp flow.  (c) Spatiotemporal velocity field resulting from a
computer simulation with the gas-kinetic-based traffic
model (GKT)~\cite{GKT}, which allows to  treat ramps easily. The arrow
indicates the driving direction.  
One can clearly see pronounced stop-and-go waves emanating from 
an area of oscillating congested traffic. 
} 
  \label{fig:rampsOffOn}
\end{figure}

We see that the instability diagram implies a large variety of
congestion patterns already in the simple simulation scenario of a
homogeneous freeway with a single ramp. The possible congestion
patterns are even richer in cases of complex freeway setups. All
combinations of the previously discussed, ``elementary'' traffic
patterns are possible. Furthermore, we expect particular patterns due
to interactions among patterns through spillover effects. For
illustration, let us focus here on the combination of an on-ramp with
an off-ramp further upstream. This freeway design is illustrated in Fig. \ref{fig:rampsOffOn} and
often built to reduce the magnitude of traffic
breakdowns, since it is favorable when vehicles leave the freeway before
new ones enter. Nevertheless, the on-ramp and the off-ramp bottleneck can get coupled,
namely when congestion upstream of the on-ramp reaches the location of the off-ramp.

What would a bottleneck analysis analogous to the one
in Sec.~\ref{sec:phasediag} predict for this setup? In order to discuss
this, let us again denote the outflow capacity downstream of the
on-ramp by $Q_{\rm out}$, its bottleneck strength equivalent to the
on-ramp flow $Q_{\rm rmp}= Q_{\rm on}$ by $\Delta Q_{\rm on} = Q_{\rm rmp}/I_{\rm fr} \ge 0$,
the upstream flow by $Q_{\rm up}$, and the average congested flow
resulting immediately upstream of the on-ramp by $Q_{\rm cong}$. In
contrast, we will denote the same quantities relating to the area of
the off-ramp by primes (${}'$), but we will introduce the abbreviation
$-\Delta Q_{\rm off} = Q'_{\rm rmp}/I_{\rm fr} \le 0$ for the effect of the
off-ramp flow $Q'_{\rm rmp}\le 0$.  
\par
According to Fig.~\ref{fig:rampsOffOn}, we observe the following dynamics: First,
traffic breaks down at the strongest bottleneck, which is the on-ramp.  If $Q_{\rm up} >
Q_{\rm out} - \Delta Q_{\rm on}$, congested flow of size $Q_{\rm cong}= Q_{\rm out} - \Delta Q_{\rm on}$ expands, and eventually reaches the
location of the off-ramp, see Fig.~\ref{fig:rampsOffOn}(a). Afterwards, the
freeway capacity downstream of the off-ramp suddenly drops from
$Q'_{\rm out}=Q_{\rm out}$ to the congested flow
\begin{equation}
 Q'_{\rm out} = Q_{\rm cong} = Q_{\rm out} - \Delta Q_{\rm on}
\end{equation} 
due to a spillover effect. This abrupt change in the bottleneck
capacity restricts the capacity for the flow {\em upstream} of the off-ramp to
\begin{equation}
Q'_{\rm cong} = Q_{\rm cong}+\Delta Q_{\rm off} \ge Q_{\rm cong} \,.  
\label{merK}
\end{equation}
 This higher flow capacity implies either free flow or milder congestion upstream of the off-ramp.
If $Q'_{\rm cong}$ is smaller
than the previous outflow capacity $Q_{\rm out}$, we have a bottleneck along the off-ramp, 
and its effective strength $\Delta Q$ is given by the {\it difference} of these values:
\begin{equation}
\Delta Q= Q_{\rm out}-(Q_{\rm cong}+\Delta Q_{\rm off})
 = \Delta Q_{\rm on} - \Delta Q_{\rm off} \, . 
\end{equation}
That is, the bottleneck strength is defined as the amount of outflow from congested traffic
which cannot be served by the off-ramp and the downstream freeway
flow. For $Q_{\rm cong}+\Delta Q_{\rm off}\ge Q_{\rm
out}$, no bottleneck occurs, which corresponds to a 
bottleneck strength $\Delta Q = 0$. This finally results in the expression
\begin{equation}
 \Delta Q =  \max(\Delta Q_{\rm on} - \Delta Q_{\rm off},0) \le \Delta Q_{\rm on}
\label{Del}
\end{equation}
\cite{helbing-sectionbased03}. Whenever $\Delta Q_{\rm off} > \Delta Q_{\rm on}$,
there is no effective bottleneck upstream of the off-ramp, i.e. the
off-ramp bottleneck is de-activated. For $\Delta Q=\Delta Q_{\rm on} -
\Delta Q_{\rm off}>0$, however, the resulting congested flow upstream
of the off-ramp becomes
\begin{equation}
 Q'_{\rm cong} = Q'_{\rm out} + \Delta Q_{\rm off} 
 = Q_{\rm out} - \Delta Q_{\rm on} +
 \Delta Q_{\rm off} = Q_{\rm out} - \Delta Q \, . 
\label{QcongOff}
\end{equation}
In conclusion, if congested traffic upstream of an on-ramp reaches an
upstream off-ramp, the off-ramp becomes a bottleneck of strength
$\Delta Q$, which is given by the difference between the on-ramp and
the off-ramp flows (or zero, if this difference would be negative).
\par
Since $\Delta Q \le \Delta Q_{\rm on}$ according to Eq.~\eqref{Del}
and $Q'_{\rm cong} \ge Q_{\rm cong}$ according to Eq.~\eqref{merK},
the congestion upstream of the off-ramp tends to be ``milder'' than
the congestion upstream of the on-ramp. The resulting traffic pattern
is often characterized by homeogeneous or oscillating congested
traffic between the off-ramp and the on-ramp, and by stop-and-go waves
upstream of the off-ramp, i.e. it has typically the appearance of a
``pinch effect''~\cite{Kerner-Rehb96-2} (see
Fig.~\ref{fig:rampsOffOn}c). For this reason, Kerner also calls the
``pinch effect'' a ``general pattern''~\cite{Kerner-book}.\footnote{Oscillatory congestion patterns upstream
of off-ramps are further promoted by a behavioral feedback, since
drivers may decide to leave the freeway in response to downstream
traffic congestion.}

\section{\label{sec:phasediagOther}Other 
Phase Diagrams and Universality Classes of Models}
\par

\begin{figure}
\centering
\includegraphics[width=0.24\textwidth]{\figpath{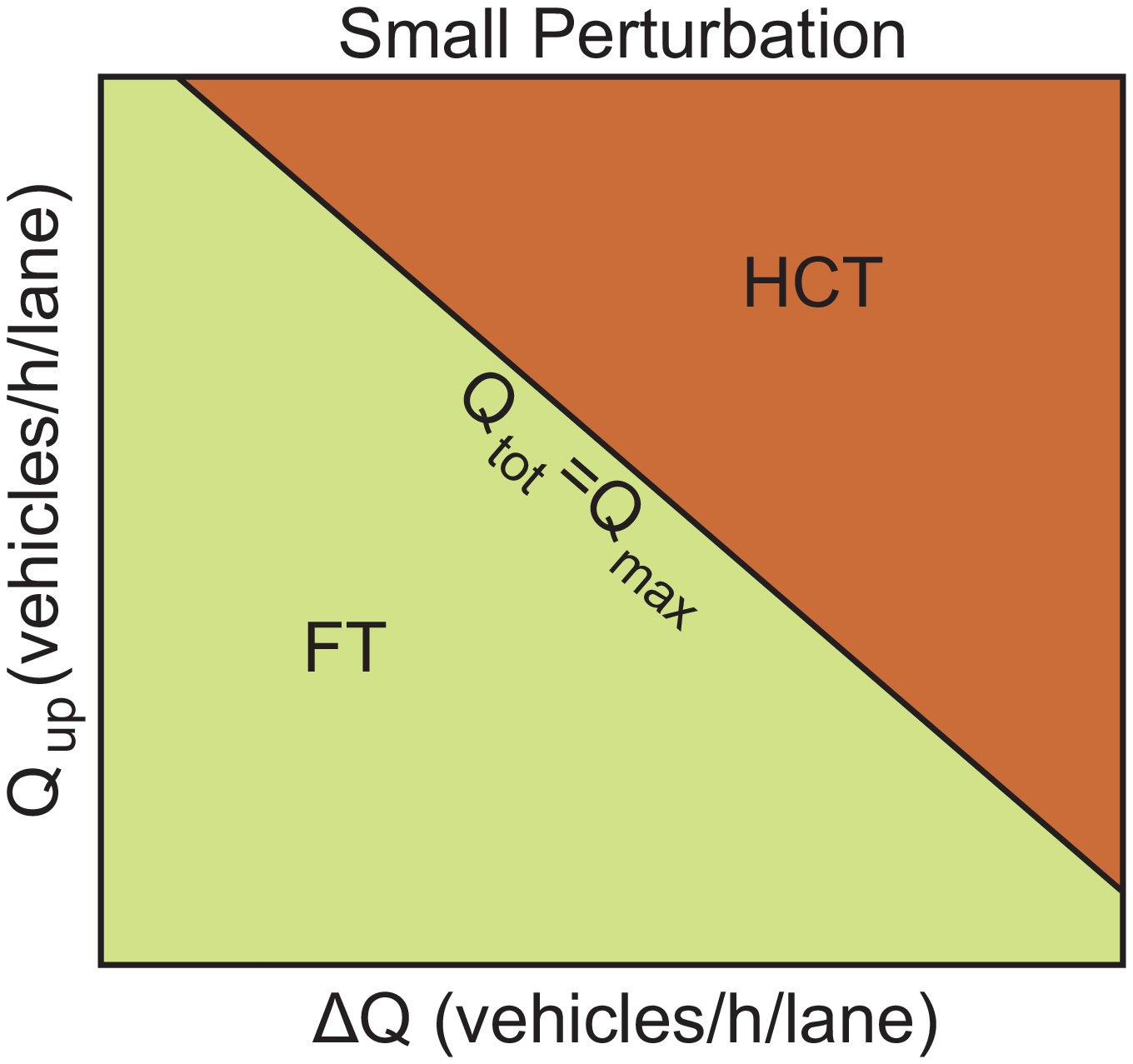}}
\includegraphics[width=0.24\textwidth]{\figpath{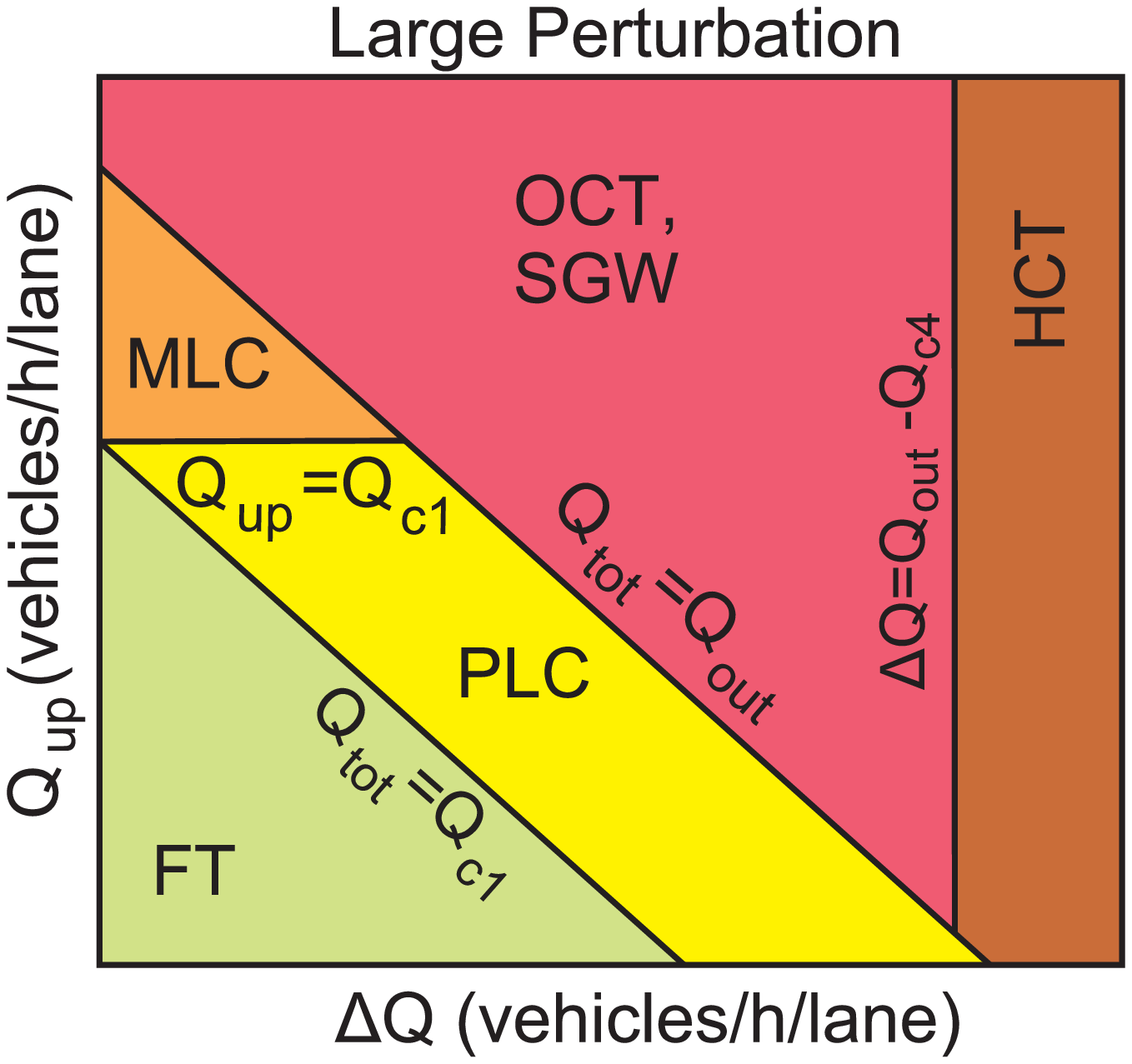}}

\caption{Schematic phase diagram for traffic flow {\it without} an extended linearly unstable density
regime ($\rho_{c2}=\rho_{c3}$), when the traffic flow at
capacity (at the density $\rhoQmax$ corresponding to the maximum flow)
is assumed to be metastable
($\rho_{c1}<\rhoQmax<\rho_{c4}$).
}

\label{fig:phasediag2c}

\end{figure}

\begin{figure}
\centering
\includegraphics[width=0.24\textwidth]{\figpath{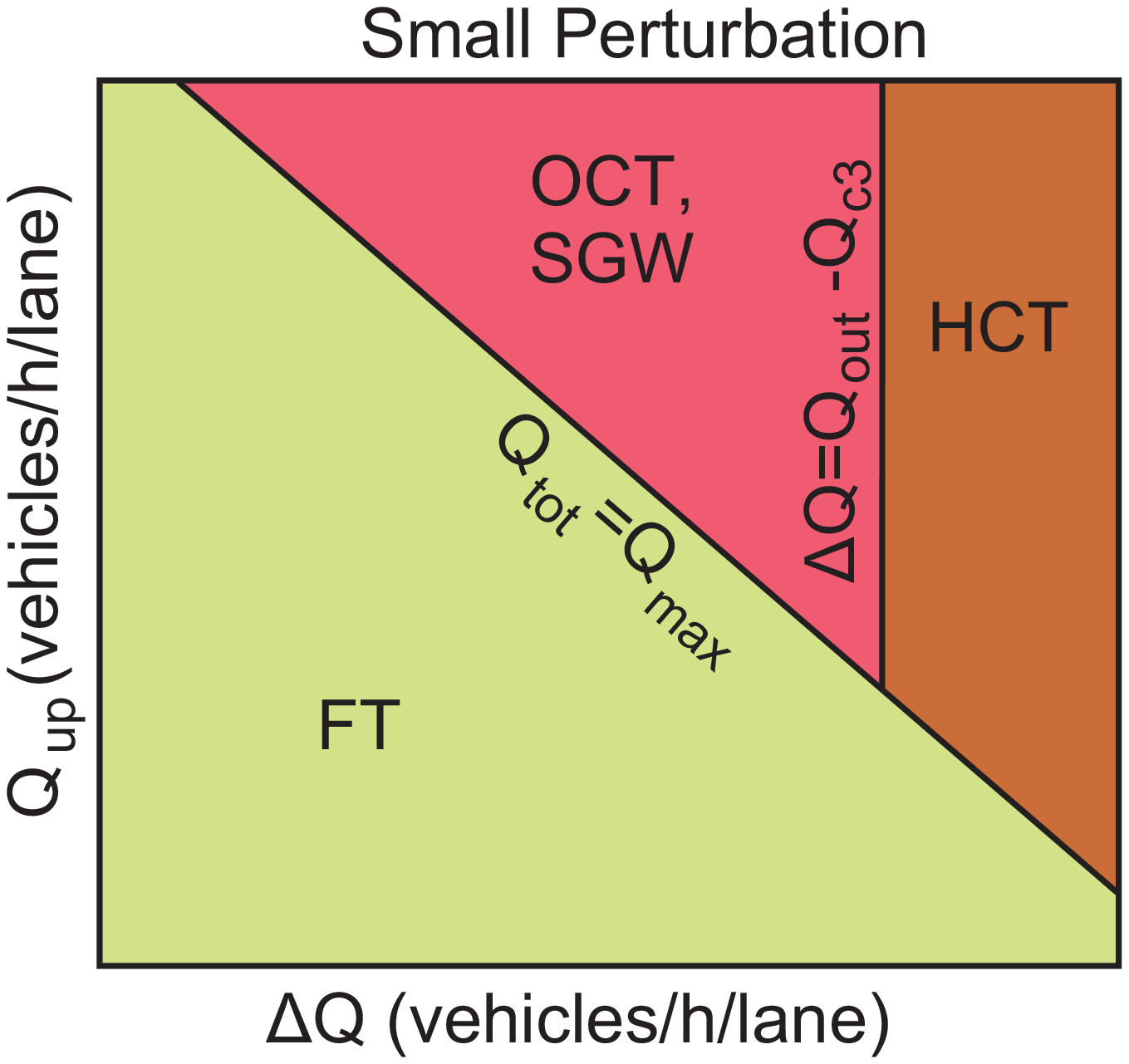}}
\includegraphics[width=0.24\textwidth]{\figpath{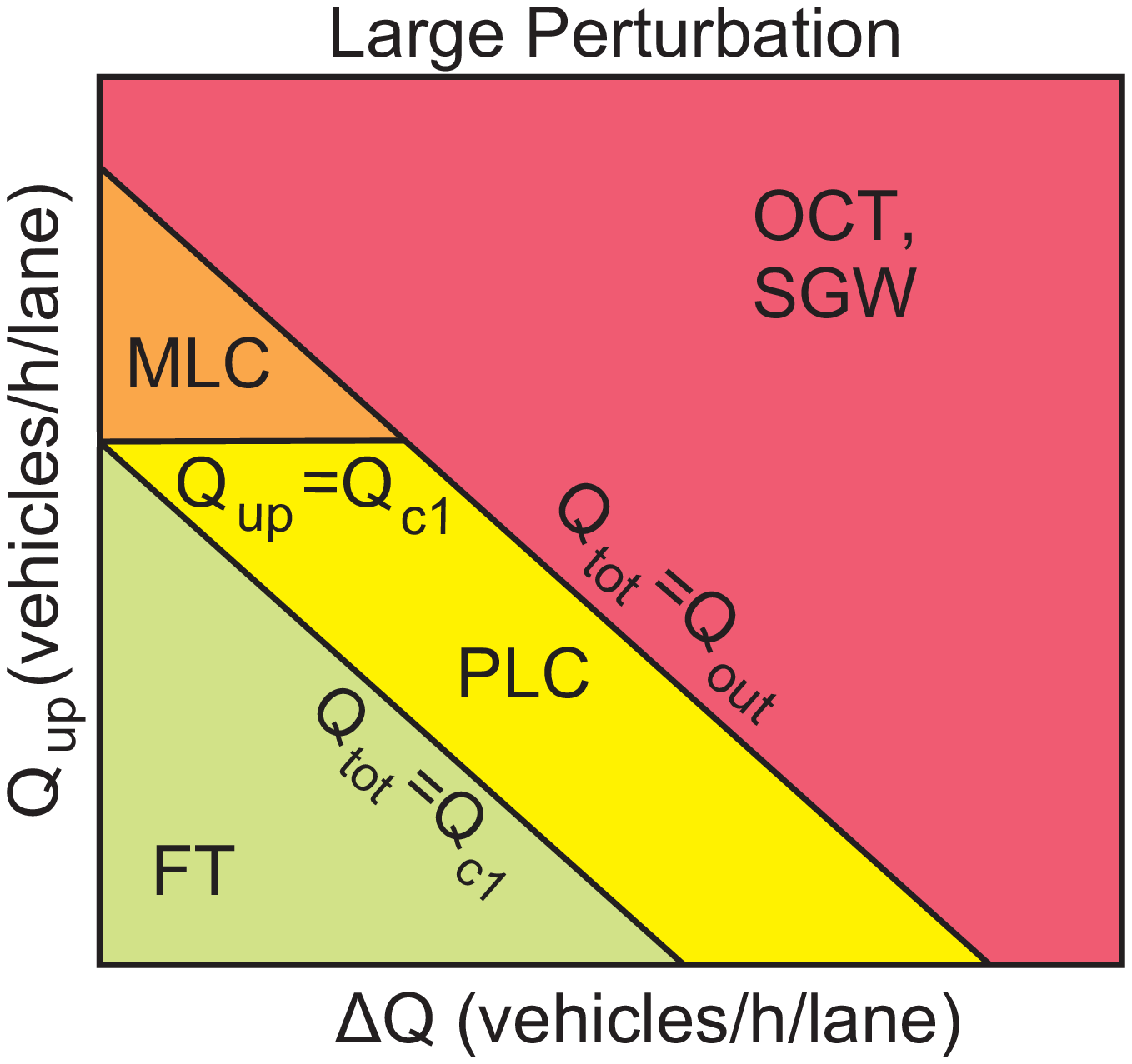}}

  \caption{Schematic phase diagram for the case of an incomplete restabilisation
at high densities, $\rho_{c3}<\rho_{c4}=\rho_\text{jam}$, when
traffic at capacity is assumed to be linearly unstable ($\rho_{c1}<\rho_{c2}<\rhoQmax$).
}
\label{fig:phasediag1e}
\end{figure}

\begin{figure}
\centering
\includegraphics[width=0.24\textwidth]{\figpath{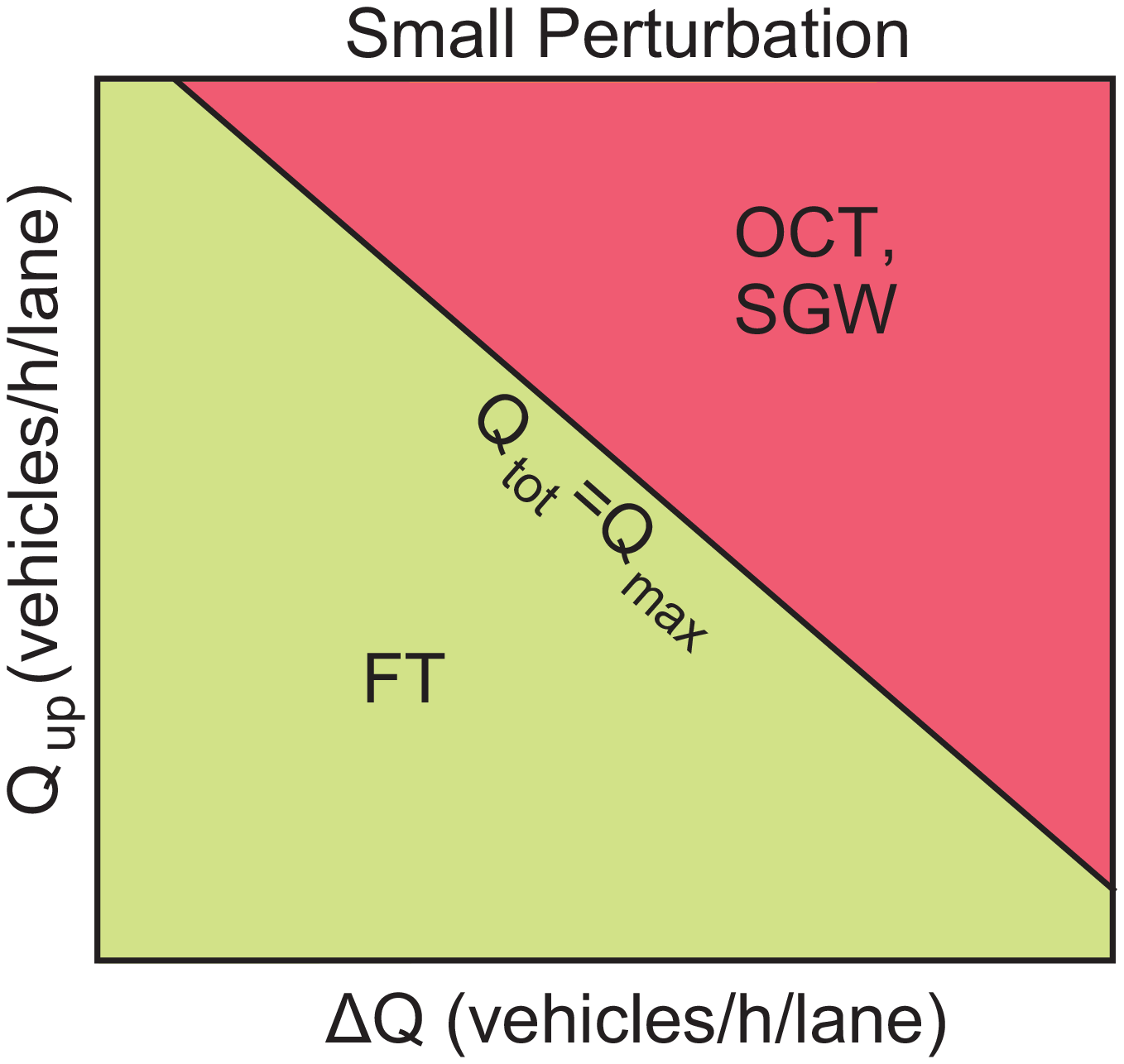}}
\includegraphics[width=0.24\textwidth]{\figpath{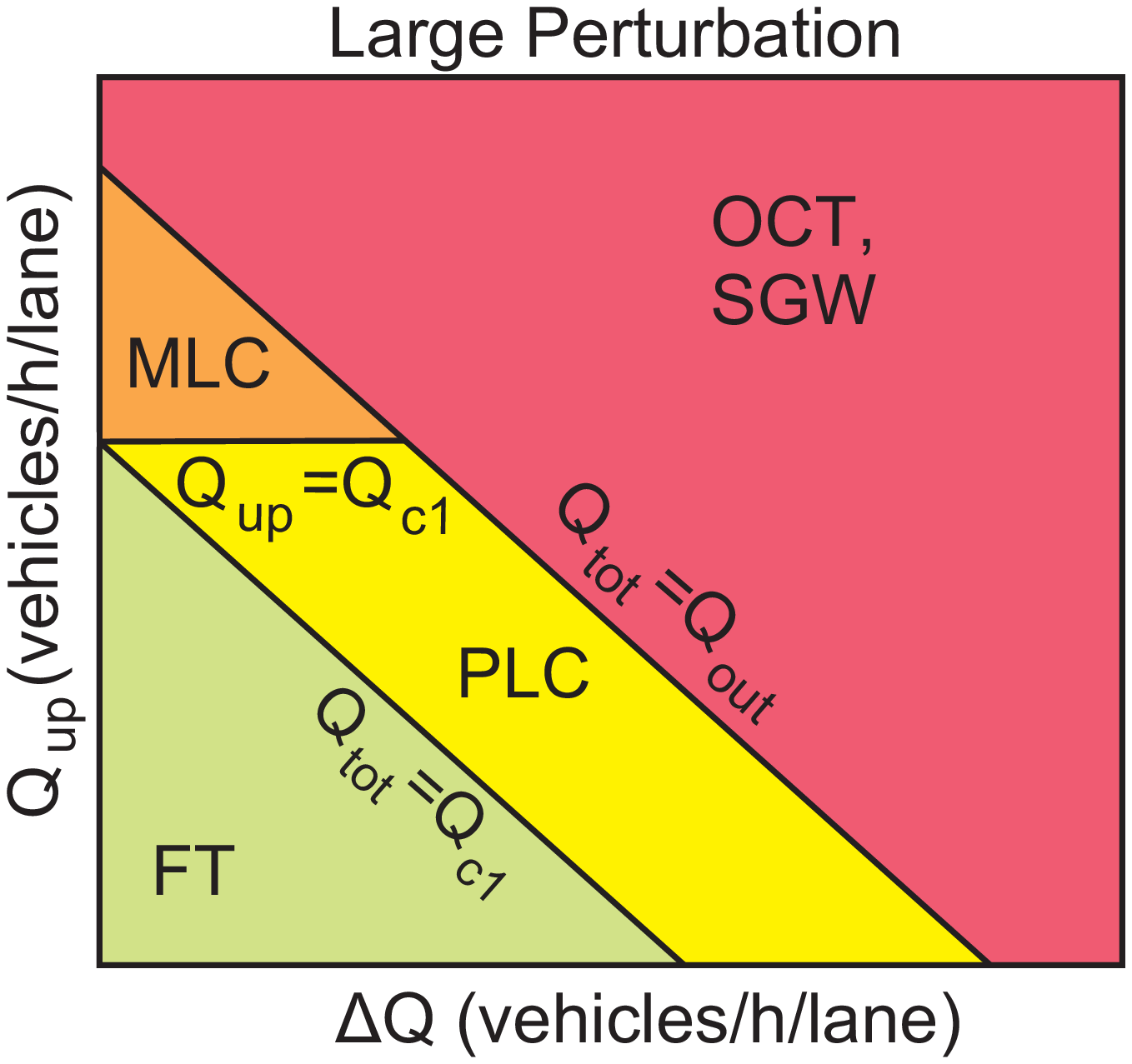}}

  \caption{Schematic phase diagram for traffic flow exhibiting both, metastable
and linearly unstable density regimes, with unstable flow at capacity
 ($\rho_{c1} < \rho_{c2} < \rhoQmax$), 
but no restabilisation for very high densities
($\rho_{c3}=\rho_{c4}=\rho_\text{jam}$).
}

\label{fig:phasediag1b}

\end{figure}

The phase diagram approach can also be used for a classification of
traffic models. By today, there are hundreds of traffic models, and
many models have a similar goodness of fit, when parameters are
calibrated to empirical
data~\cite{HelbTilch-98ab,Brockfeld-benchmark03,Brockfeld-benchmark04,Ossen-benchmark05,Ossen-interDriver06,Kesting-Calibration-TRR08}. It
is, therefore, difficult, if not impossible, to determine ``the best''
traffic model. However, one can classify models according to
topologically equivalent phase diagrams. Usually, there would be
several models in the same universality class, producing qualitatively
the same set of traffic patterns under roughly similar
conditions. Among the models belonging to the same universality class,
one could basically select {\it any} model. According to the above,
the differences in the goodness of fit are usually not
dramatic. Models with many model parameters may even suffer from
insignificant parameters or parameters, which are hard to calibrate,
at the cost of predictive power. Therefore, it is most reasonable to
choose the {\it simplest} representative of a universality class
which, however, should fulfil minimum requirements regarding
theoretical consistency.
\par
Before we enter the comparison with empirical data, let us discuss a
number of phase diagrams expected for certain kinds of traffic models.
Particular specifications of the optimal velocity
model, for example, are linearly unstable
for one density $\rho_{\rm c2} = \rho_{\rm c3}$ only, but show unstable behavior in an extended density regime for sufficiently large perturbations (i.e. extended metastable regimes)~\cite{HelbMoussaid-EPJB-08}. The schematic phase diagram expected in this
case is shown in Fig.~\ref{fig:phasediag2c}. Some other traffic models
have linearly unstable and metastable regimes, but do not show a
restabilisation at very high densities, i.e. $\rho_{\rm
c4}=\rho_\text{jam}$ (see Fig.~\ref{fig:phasediag1e}), and sometimes
one even has $\rho_{\rm c3}=\rho_\text{jam}$~\cite{siebel2006,KKW-CA}
(see Fig. \ref{fig:phasediag1b}). In the latter case, homogeneous
congested traffic does not exist. In models such as the IDM, the
restabilisation depends on the chosen parameter
values~\cite{Treiber-ThreePhasesTRB}, see also
Appendix~\ref{app:IDMstab}.

\par
In most of the currently studied traffic models, one has either {\it both}, linearly unstable
{\it and} metastable density ranges, or unconditionally stable traffic. In
principle, however, models with linearly unstable but no metastable
regimes are conceivable. For example, they may be established by taking 
a conventional model and introducing a dependence on
the square of the velocity gradient 
(in macroscopic models) or the velocity difference (in microscopic models).
\par
\begin{figure}
\centering
\includegraphics[width=0.24\textwidth]{\figpath{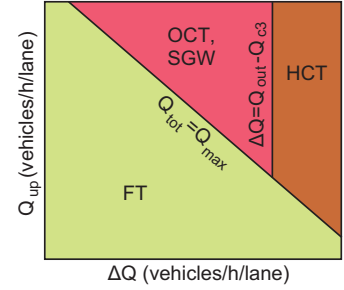}}

  \caption{Schematic phase diagram, if there are only stable and linearly
unstable, but no metastable density regimes ($\rho_{c1}=\rho_{c2}$,
$\rho_{c3}=\rho_{c4}$). Furthermore, traffic at capacity is assumed to
be unstable 
($\rho_{c1}<\rhoQmax<\rho_{c4}$)
}

\label{fig:phasediag1d}

\end{figure}

 
A linearly unstable model without metastable ranges would correspond to
$\rho_{\rm c2} = \rho_{\rm c1}$ and $\rho_{\rm c4} = \rho_{\rm
c3}$. For such models, we do not expect any multi-stability (see Fig.~\ref{fig:phasediag1d}), 
and localized congested traffic would only be possible under special
conditions \cite{Lee-emp,Martin-empStates} (e.g. on freeway sections between off- and on-ramps). If, in addition,
there is no restabilisation  (i.e. $\rho_{\rm c3}=\rho_\text{jam}$),
only free traffic and oscillating congested traffic should exist. This seems to reflect
the situation for the  classical Nagel-Schreckenberg
model~\cite{Nagel-S}, although the situation is somewhat unclear, since
this model is stochastic and an exact distinction between free and
congested states is difficult in this model.

\par
\begin{figure}
\centering
\includegraphics[width=0.24\textwidth]{\figpath{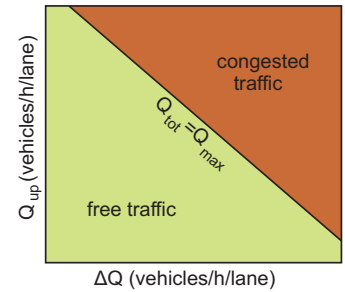}}

  \caption{Schematic phase diagram, if traffic is unconditionally
  stable ($\rho_{c1}=\rho_{c2}=\rho_{c3}=\rho_{c4}$). The most
  prominent example is the Lighthill-Whitham model \cite{Lighthill-W},
  but many other models (including the gas-kinetic-based traffic model
  (GKT)~\cite{GKT} and the IDM) can be parameterized to reproduce this
  case.}

\label{fig:phasediag3}
\end{figure}

Finally, we would like to discuss the fluid-dynamic model by Lighthill
and Whitham~\cite{Lighthill-W}, which does not display {\it any}
instabilities ~\cite{Helb-Opus} and, consequently, has only
homogeneous patterns, namely free traffic
for $Q_{\rm tot} \le Q_{\rm max}$ and (homogeneous) extended congested
traffic for $Q_{\rm tot} > Q_{\rm max}$ (which corresponds to a
vehicle platoon behind the bottleneck). This is illustrated in
Fig.~\ref{fig:phasediag3}. The two phases can also be
distinguished locally, if temporal correlations are considered: While
perturbations in free traffic travel in forward direction, in the
congested regime they travel backward.

\par
We underline again that, by changing model parameters (corresponding to different
driving styles), the resulting instability and phase diagrams of many
traffic models change as well. For example, the IDM can be
parameterized to generate most of the stability diagrams discussed
in this contribution.  Since different parameter values correspond to
different driving styles or prevailing velocities, this may explain differences between
empirical observations in different countries. For example,
oscillating congested traffic seems to occur less frequent in the
United States~\cite{CasBer-99,Zielke-intlComparison}. 
\par
 Finally, note that somewhat different phase diagrams result for models that are 
characterized by a complex vehicle dynamics and no existence of a
fundamental diagram~\cite{KKW-CA,Kerner-PhysA-04}. Nevertheless, similarities can be discovered
(see Sec.~\ref{sec:phasediag}).

\section{\label{sec:phasediagEmp}Empirical Phase Diagram}

\begin{figure}[htbp]
\centering
 \includegraphics[width=6.5cm]{\figpath{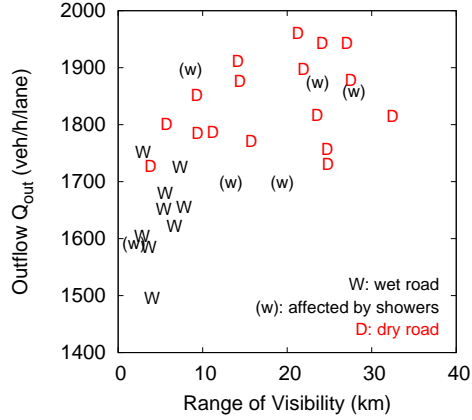}}
  \caption{The outflows $Q_{\rm out}$ of
congestion patterns correlate significantly with the weather-dependent
range of visibility (meteorological optical range $3/\sigma$). This
was determined by 
measuring the extinction coefficient $\sigma$, using multiple laser
reflections. ``W'' denotes traffic breakdowns when the road surface was wet, while
``D'' stands for a dry road surface. Note that there are three cases
of congestion marked with a lower-case ``w'', which were related with
a short and light shower only, so that the outflow values stayed
comparatively high. For details see Ref.~\cite{Schoenhof-Diss}.
}
\label{fig:outflow_visibility}
\end{figure}

\begin{figure}
\centering

\includegraphics[width=6.5cm]{\figpath{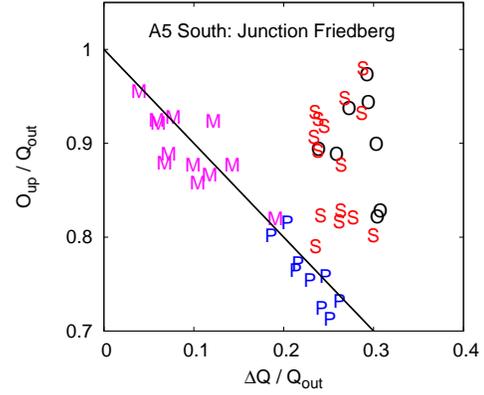}}

 \caption{
 Empirical phase diagram, where the flows have been
 scaled by the respective outflows $Q_{\rm out}$
 (after~\cite{Martin-empStates}). The data represent the congested
 traffic states observed on the German freeway A5 at Junction
 Friedberg in direction South (M = moving localized
 cluster, S = stop-and-go waves, O = oscillating congested traffic, P
 = pinned localized cluster). It can be clearly seen that the
 non-extended traffic states are scattered around the line $Q_{\rm tot}/Q_{\rm out} = (Q_{\rm
 up}+ \Delta Q )/ Q_{\rm out} = 1$, as expected, while the
 extended traffic states are above this line.  Moreover, pinned
 localized clusters, moving localized clusters, and stop-and-go
 waves/oscillating congested traffic  are well separated from each other.
 Homogeneous congested traffic, but not other traffic states were observed for 
 $\Delta Q/Q_{\rm out} \gtrsim 0.5$ (see Ref.~\cite{Martin-empStates}). }
\label{fig:empirical_PD}
\end{figure}

The remaining challenge in this paper is to find the universality
class that fits the stylized facts of traffic dynamics well. Here, we
will primarily demand that it fits the empirical phase diagram, i.e.
reproduces all elementary congestion patterns observed, and not
more. We have evaluated empirical data from the German freeway A5
close to Frankfurt. Due to the weather-dependence of the outflows $Q_{\rm out}$
(see Fig. \ref{fig:outflow_visibility}), it is important to scale all flows by
the respective measurements of $Q_{\rm out}$. This naturally collapses the area of 
localized congested traffic states to a line. As
Fig.~\ref{fig:empirical_PD} shows, the phase diagram after scaling the
flows is very well compatible with the theoretical phase
diagrams of Figs.~\ref{fig:phasediag1a} and~\ref{fig:phasediag2a}. 
Since the determination of the empirical phase diagram did not focus on the
detection of ``widening synchronized patterns'', it does not allow us to clearly
distinguish between the two phase diagrams, i.e. to decide whether
$\rho_{\rm c2} > \rhoQmax$ or $\rho_{\rm c2} < \rhoQmax$. However, the empirical
WSP displayed in Fig. \ref{fig:elemPatterns} suggests that Fig. \ref{fig:phasediag2a}
corresponding to $\rho_{\rm c2} > \rhoQmax$  would be the right choice.
Another piece of evidence for this is the metastability of vehicle platoons forming behind overtaking
trucks (see Ref.~\cite{HelbTilch-EPJB-08}).\footnote{The existence of ``widening moving clusters'', see Sec.~\ref{sec:classicalPatterns} and Fig.~\ref{fig:elemPatterns}(a), supports this view as well.}

\subsection{\label{sec:reply}Reply 
to Criticisms of Phase Diagrams for Traffic Models with a Fundamental Diagram}\label{Kerner1}

In the following, we will face the criticism of the phase diagram approach 
by Kerner \cite{BKerner-98,Kerner-book}:
\begin{enumerate}
\item Models containing a fundamental diagram could not explain the
wide scattering of flow-density data observed for ``synchronized''
congested traffic flow. This is definitely wrong, as a wide scattering
is excellently reproduced by considering the wide distribution of
vehicle gaps, partially due to different vehicle classes such as cars and trucks~\cite{GKT-scatter,Katsu03}. Note that, for a good reproduction of empirical measurements,
it is important to apply the same measurement procedure to empirical and simulated data, 
in particular the data aggregation over a finite time period.
\item As the ramp flow or the overall traffic volume $Q_{\rm tot}$ is increasing,
the phase diagram approach would predict the transitions free traffic
$\rightarrow$ moving or pinned localized cluster $\rightarrow$
stop-and-go traffic/oscillating congested traffic $\rightarrow$
homogeneous congested traffic. However, this would be wrong because
(i) homogeneous congested traffic would not exist~\cite{Kerner-00,Ker02_PRE}, and 
(ii) according to the ``pinch effect'', wide moving jams (i.e. moving localized clusters) should occur {\it after} the occurrence of ``synchronized flow'' (i.e. extended congested traffic) \cite{Kerner-98e}.
\par
We reply to (i) that it would be easy to build a car-following model with a fundamental diagram that produces no HCT states,\footnote{An example would be the IDM with the parameter choice $s_1 =0$.} but according to empirical data, homogeneous congested traffic 
{\it does} exist (see Fig.~\ref{fig:elemPatterns}f), but it occurs very rarely and only
for extremely large bottleneck strengths exceeding $\Delta Q \approx 0.5\, Q_{\rm out}$  
\cite{Martin-empStates}. As freeways are dimensioned such that
bottlenecks of this size are avoided, HCT occurs primarily when freeway lanes are
closed after a serious accident.  In other words, when excluding cases
of accidents from the data set, HCT states will normally not be
found. 
\par
Moreover, addressing point (ii), Kerner is wrong in claiming that our theoretical
phase diagram would necessarily {\it require} moving localized clusters to occur {\it
before} the transition to stop-and-go waves or oscillating congested
traffic. This misunderstanding might have occurred by ignoring the
dependence of the resulting traffic state on the perturbation size. The OCT pattern of 
Fig.~\ref{fig:elemPatternsIDM}(c) clearly shows that a
{\it direct} transition from free traffic flow to oscillating congested
traffic is possible in cases of small perturbations. The same applies to a fast increase
in the traffic volume $Q_{\rm tot}(t)$, which is typical during rush hours.
\item  The variability of the empirical outflow $Q_{\rm out}$ would
not be realistically 
accounted for by traffic models with a fundamental diagram. This variability, however, does not require an 
explanation based on complex vehicle dynamics. To a large extent, it can be understood by variations
in the weather conditions (see Fig. \ref{fig:outflow_visibility}) and
in the flow conditions on the freeway lanes in the neighborhood of
ramps \cite{Martin-empStates}, which is particularly affected by a
largely varying truck fraction \cite{GKT-scatter}.   
\end{enumerate}
In summary, the phase diagram approach for traffic models with a
fundamental diagram has been criticized with invalid arguments.

\subsection{On the Validity of Traffic Models}

In the past decades, researchers have proposed a large number of traffic models and it seems that
there often exist several different explanations for the same observation(s)~\cite{Treiber-ThreePhasesTRB}. As a consequence,
it is conceivable that there are models which are macroscopically correct (in terms of reproducing the observed congestion patterns
discussed above), but microscopically wrong. In order to judge the validity of competing traffic models, we
consider it necessary to compare models in a {\it quantitative} way, based on empirical data. This should include
\begin{itemize}
\item a definition of suitable performance measures (such as the deviation between simulated and
measured travel times or velocity profiles), 
\item the implementation and parameter calibration of the competing
models with {\it typical} empirical data sets, and
\item the comparison of the performance of the competing models for
different {\it test} data sets of representative traffic situations.
\end{itemize}
Based on data sets of car-following experiments, such analyses have,
for example, been performed with a number of follow-the-leader
models~\cite{HelbTilch-98ab,Brockfeld-benchmark03,Brockfeld-benchmark04,Ossen-benchmark05,Ossen-interDriver06,Kesting-Calibration-TRR08},
with good results in particular for the intelligent driver model
(IDM)~\cite{Brockfeld-benchmark03,Brockfeld-benchmark04,Kesting-Calibration-TRR08}.
If there is no statistically significant difference in the performance
of two models (based on an analysis of variance), preference should be
given to the simpler one, according to Einstein's principle that a
model should be always as simple as possible, but not simpler.
\par
We would like to point out that over-fitting of a model must be avoided. This may easily happen for models with many parameters.
Fitting such models to data will, of course, tend to yield smaller errors than fitting models with a few parameters
only. Therefore, one needs to make a significance analysis of parameters that {\it adjusts} for the number of parameters, as it is commonly done in statistical analyses. Reproducing a certain {\it calibration} data set well does not necessarily mean that an independent {\it test} data set will be well reproduced. While the {\it descriptive} capability of models with many parameters is often high, models with fewer parameters may have a higher {\it predictive} capability, as their parameters are often easier to calibrate.
\par
This point is particularly important, since it is known that traffic
flows fluctuate considerably, especially in the congested regime. So,
one may pose the question whether these fluctuations are meaningful
dynamical features of traffic flows or just noise. To some extent,
this depends on the question to be addressed by the model, i.e. how
fine-grained predictions the model shall be able to make. There are
certainly systematic sources of fluctuations, such as lane-changes, in
particular by vehicles entering or leaving the freeway via ramps,
different types of vehicles, and different driver
behaviors~\cite{Ossen-interDriver06,Kesting-Calibration-TRR08}. Such issues would most naturally
be addressed by multi-lane models considering lane changes and
heterogeneous driver-vehicle units~\cite{MOBIL-TRR07}. Details like
this may, in particular, influence the outflow $Q_{\rm out}$ of
congested traffic flow (see Fig. 17 in
Ref. \cite{Martin-empStates}). When trying to understand the
empirically observed variability of the outflow, however, one also
needs to take the variability of the {\it weather conditions} and the
visibility into account (see Fig. \ref{fig:outflow_visibility}).  In
order to show that car-following models with a fundamental diagram are
inferior to other traffic models in terms of reproducing microscopic
features of traffic flows (even when multi-lane multi-class features
are considered), one would have to show with standard statistical
procedures that these other models can explain a larger share of the
empirically observed variance, and that the difference in the
explanation power is significant, even when the number of model
parameters is considered. To the knowledge of the authors, however,
such a statistical analysis has not been presented so far.

\section{\label{sec:conclusion}Summary, Conclusions, and Outlook}

After a careful discussion of the term ``traffic phase'', we have
extended the phase diagram concept to  traffic models with a
fundamental diagram  that are not only capable of reproducing
congestion patterns such as localized clusters, stop-and-go waves,
oscillatory congested traffic, or homogeneous congested traffic, but
also ``widening 
synchronized patterns'' (WSP) and ``widening moving clusters''
(WMC). The discovery of these states for the case, where the maximum
traffic flow lies in the metastable density regime, was quite
unexpected. It offers an alternative and -- from our point of
view -- simpler interpretation of some of Kerner's empirical
findings. A particular advantage of starting from models with a
fundamental diagram is the possibility of {\it analytically} deriving
the schematic phase diagram of traffic states from the instability
diagram, which makes the approach {\it predictive}.
\par
Furthermore, we have discussed how the
phase diagram approach can be used to classify models into
universality classes. Models within one universality class are
essentially equivalent, and one may choose any, preferably the simplest
representative satisfying minimum requirements regarding theoretical
consistency. The universality class should be chosen in agreement with
empirical data.  These were well represented by the schematic phase diagram 
in Fig. \ref{fig:phasediag2a}. Furthermore, we have
demonstrated that one needs to implement the full details of a
freeway design, in particularly all on- and off-ramps, as these
details matter for the resulting congestion patterns. Multi-ramp
designs lead to congestion patterns composed of several elementary congestion
patterns, but spillover effects must be considered. In this way, a
simple explanation of the ``pinch effect''~\cite{Kerner-98e} and the
so-called ``general pattern''~\cite{Kerner-book} results. We have also replied to
misunderstandings of the phase diagram concept. 
\par
In conclusion, the phase diagram approach is a simple and natural
approach, which can explain empirical findings well, in particular the
dependence of traffic patterns on the flow conditions. Note that the
phase diagram approach is a metatheory rather than a model. It can be
theoretically derived from the instability diagram of traffic flows
and the self-organized outflow from seriously congested traffic. This
is not a triviality and, apart from this, the phase diagram approach
is more powerful than the instability diagram itself: It does not only
allow predictions regarding the possible appearance of traffic
patterns and possible transitions between them. It also allows to
predict whether it is an extended or localized traffic pattern, or
whether a localized cluster moves or not. Furthermore, it facilitates
the prediction of the spreading dynamics of congestion in space, as
reflected by Eqs.~\eqref{C1} and \eqref{C2}. This additionally
requires formula~\eqref{Qcong}, which determines how the bottleneck
strength $\Delta Q$ determines the effective flow capacity $Q_{\rm cong}$
of the upstream freeway section.

\begin{acknowledgement}
DH and MT are grateful for the inspiring discussions with the
participants of the Workshop on ``Multiscale Problems and Models in
Traffic Flow'' organized by Michel Rascle and Christian Schmeiser at
the Wolfgang Pauli Institute in Vienna from May 5--9, 2008, with
partial support by the CNRS. Furthermore, the authors would like to
thank for financial support by the Volkswagen AG within the BMBF
research initiative INVENT and the {\it Hessisches Landesamt f\"ur
Stra{\ss}en und Verkehrswesen} for providing the freeway data.
\end{acknowledgement}



\begin{appendix}
\section{\label{app:SourceSink}Modeling of Source and Sink Terms (In- and Outflows)}

In this appendix, we will focus on the case of 
a freeway section with a single bottleneck such as an isolated on-ramp. Scenarios with several bottlenecks are discussed in Sec. \ref{sec:rampsOffOn}. 
\par
In order to derive the appropriate form of source and sink terms due to on- or off-ramps, we start from the continuity equation, which reflects the conservation of the number of vehicles. If $\rho_*(x,t)$ represents the one-dimensional density of vehicles at time $t$ and a location $x$ along the freeway, and if $Q_*(x,t)$ represents the vehicle flow measured at a cross section of the freeway, the continuity equation can be written as follows:
\begin{equation}
\frac{\partial \rho_*(x,t)}{\partial t} + \frac{\partial Q_*(x,t)}{\partial x} = 0 \, .
\label{conti}
\end{equation}
Now, assume that $I(x)$ is the number of freeway lanes at location $x$. We are interested in 
the density $\rho(x,t) = \rho_*(x,t)/I(x)$ and traffic flow $Q(x,t)= Q_*(x,t)/I(x)$ per freeway lane.
Inserting this into the continuity equation (\ref{conti}) and carrying out partial differentiation, applying the product rule of Calculus, we get
\begin{eqnarray}
\frac{\partial}{\partial t} \big[ I(x) \rho(x,t)\big] &=&
 I(x) \frac{\partial \rho(x,t)}{\partial t} \nonumber \\
 &=& - \frac{\partial}{\partial x} \Big[  I(x) Q(x,t) \Big] \nonumber \\
  &=& - Q(x,t) \frac{d I(x)}{d x} - I(x) \frac{\partial Q(x,t)}{\partial x} \, . \qquad
\label{right}
\end{eqnarray}
Rearranging the different terms, we find
\begin{equation}
 \frac{\partial \rho(x,t)}{\partial t} + \frac{\partial Q(x,t)}{\partial x}
  = - \frac{Q(x,t)}{I(x)} \frac{d I(x)}{d x }  \, .  
  \label{withchange}
\end{equation}
The first term of this equation looks exactly like the continuity
equation for the density $\rho_*(x,t)$ over the whole cross section at
$x$. The term on the right-hand side of the equality sign describes an
increase of the density $\rho(x,t)$ per lane, whenever  the number
of freeway lanes is reduced ($\partial I(x)/\partial x < 0$)  and all
vehicles have to squeeze into the remaining lanes. In contrast, the
density per lane $\rho(x,t)$ goes down, if the width of the road
increases ($\partial I(x)/\partial x > 0$).  
\par
It is natural to treat on- and off-ramps in a similar way by the continuity equation
\begin{equation}
\frac{\partial \rho(x,t)}{\partial t} + \frac{\partial Q(x,t)}{\partial x} = \nu_{_+}(x,t) - \nu_{_-}(x,t)  
\label{conti2}
\end{equation}
with source terms $\nu_{_+}(x,t)$ and sink terms $-\nu_{_-}(x,t)$.
For example, if a one-lane on-ramp flow $Q_{\rm on}(t)$ is entering the freeway 
uniformly over an effectively used ramp length of $L_{\rm eff}$, we have $dI(x)/dx = 1/L_{\rm eff}$, which together with (\ref{right}) and~(\ref{conti2}) implies
\begin{equation}
\nu_{_+}(x,t) = \left\{
\begin{array}{ll}
\displaystyle \frac{Q_{\rm on}(t)}{I_{\rm fr} L_{\rm eff}} &\mbox{for } x_{\rm rmp} - \frac{L_{\rm eff}}{2} < x < x_{\rm rmp} + \frac{L_{\rm eff}}{2}, \\ 
0 & \mbox{otherwise.}
\end{array}
\right.
\end{equation} 
$I_{\rm fr} = I(x_{\rm rmp} \pm L_{\rm eff}/2)$ denotes the number of freeway lanes {\it upstream} and {\it downstream} of the ramp, which is assumed to be the same, here. The sink term due to off-ramp flows $Q_{\rm off}(t)\ge 0$ has the form 
\begin{equation}
\nu_-(x,t) = \left\{
\begin{array}{ll}
\displaystyle \frac{Q_{\rm off}(t)}{I_{\rm fr}L_{\rm eff}} &\mbox{for } x_{\rm rmp} - \frac{L_{\rm eff}}{2} < x < x_{\rm rmp} + \frac{L_{\rm eff}}{2} \, , \\
0 & \mbox{otherwise.}
\end{array}
\right.
\end{equation}

\section{\label{app:IDMstab}Parameter Dependence of the Instability Thresholds in the Intelligent Driver Model}

The acceleration function $a\sub{IDM}(s,v,\Delta v)$ of the
intelligent driver model (IDM)~\cite{Opus} depends on the gap $s$ to
the leading vehicle, the velocity $v$, and the velocity difference
$\Delta v$ (positive, when approaching). It is given by
\be
a\sub{IDM}(s,v,\Delta
v)=a\left[1-\left(\frac{v}{v_0}\right)^4 - \left(\frac{s^*(v,\Delta
v)}{s}\right)^2\right],
\ee
where
\be
s^*(v,\Delta v)=s_0+s_1\sqrt{\frac{v}{v_0}} + T v+\frac{v \, \Delta
v}{2\sqrt{ab}}.
\ee
For identical driver-vehicle units, there exists a one-parameter class
of homogeneous and stationary solutions defining the ``microscopic''
fundamental diagram $v_e(s)$ \textit{via}
$a\sub{IDM}(s,v_e(s),0)=0$. From a standard linear analysis around this
solutions it follows that the IDM is linearly stable if the condition
\be
\label{linstabIDM}
\ablpart{a\sub{IDM}}{s} \le \ablpart{a\sub{IDM}}{v} 
\left(\ablpart{a\sub{IDM}}{\Delta v} +\frac{1}{2}\ablpart{a\sub{IDM}}{v} 
\right)
\ee
is fulfilled. With the micro-macro relation
\be
s=\frac{1}{\rho}-l\sub{veh}, \quad \mbox{where }l\sub{veh}=\unit[6]{m},
\ee
this defines the stability boundaries $\rho\sub{c2}$ and
$\rho\sub{c3}$ as a function of the model parameters 
$v_0$ (desired velocity),
$T$ (desired time headway),
$a$ (desired acceleration), 
$b$ (desired deceleration),
$s_0$ (minimum gap), and
$s_1$ (gap parameter; if nonzero, the fundamental diagram has an
inflection point).
The overall stability can be controlled most effectively by the
acceleration $a$. Setting the other parameters to the values used in 
Fig. \ref{fig:stabdiag} [$v_0=\unit[128]{km/h}$,
$T=\unit[1]{s}$,
$s_0=\unit[2]{m}$,
$s_1=\unit[10]{m}$, and
$b=\unit[1.3]{m/s^2}$], we obtain
\bi
\item unconditional linear stability for $a \ge
\unit[1.68]{m/s^2}$,
\item linear instability in the density range $\rho_{c2} \le \rho \le 
\rho_{c3}$ for $\unit[0.95]{m/s^2} \le a \le \unit[1.68]{m/s^2}$,
 where $\rho_{c2}>\rho\sub{max}$ and $\rho_{c3}<\rho\sub{jam}$.
 In this situation, corresponding to Figs. \ref{fig:stabdiag}(c)
and (d), the instability range lies completely on the ``congested''
side of the fundamental diagram.
\item Finally, for
$a \le \unit[0.95]{m/s^2}$, the linear instability also extends to the
``free branch'' of the fundamental diagram
($\rho_{c2}<\rho\sub{max}$), corresponding to Figs.
\ref{fig:stabdiag} (a) and (b).
\ei

The upper instability threshold $\rho_{c3}$ can be controlled nearly
independently from the lower instability threshold $\rho_{c2}$ by the
gap parameters $s_0$ and $s_1$. Generally, 
$\rho_{c3}$ increases with decreasing values of $s_1$. 
In particular, if $s_1=0$, one
obtains the analytical result $\rho_{c3}=(l\sub{veh}+s_0)^{-1}$
for any $a<s_0/T^2$, and unconditional linear stability for $a>s_0/T^2$. 
As can be seen from the last expression, the instability generally
becomes more pronounced 
for decreasing values of the time headway parameter $T$, which is
plausible. 

The additional influence of the parameter $b$ according to computer
simulations is plausible as well: With decreasing values of $b$, the
sensitivity with respect to velocity differences increases, and the
instability tends to decrease.  Further simulations suggest that the
IDM has metastable density areas only when linearly unstable densities
exist. Metastability at densities above the linear instability range
additonally requires $s_1>0$.

\end{appendix}
\end{document}